\documentclass[a4paper,11pt]{article}
\usepackage{jinstpub} % for details on the use of the package, please see the JINST-author-manual
\usepackage{lineno}
\usepackage{multirow}
\usepackage{caption}
\usepackage{subcaption}
\usepackage{textcomp}
\usepackage[utf8]{inputenc}
\usepackage{graphicx}
\usepackage{siunitx}
\usepackage{url}
\usepackage[export]{adjustbox}
\usepackage{float}
\usepackage{placeins}
\usepackage{gensymb}
\usepackage[table]{xcolor}

\title{The beamformed trigger of RNO-G: \\
its design and in-field performance }

\collaboration{\includegraphics[height=19mm]{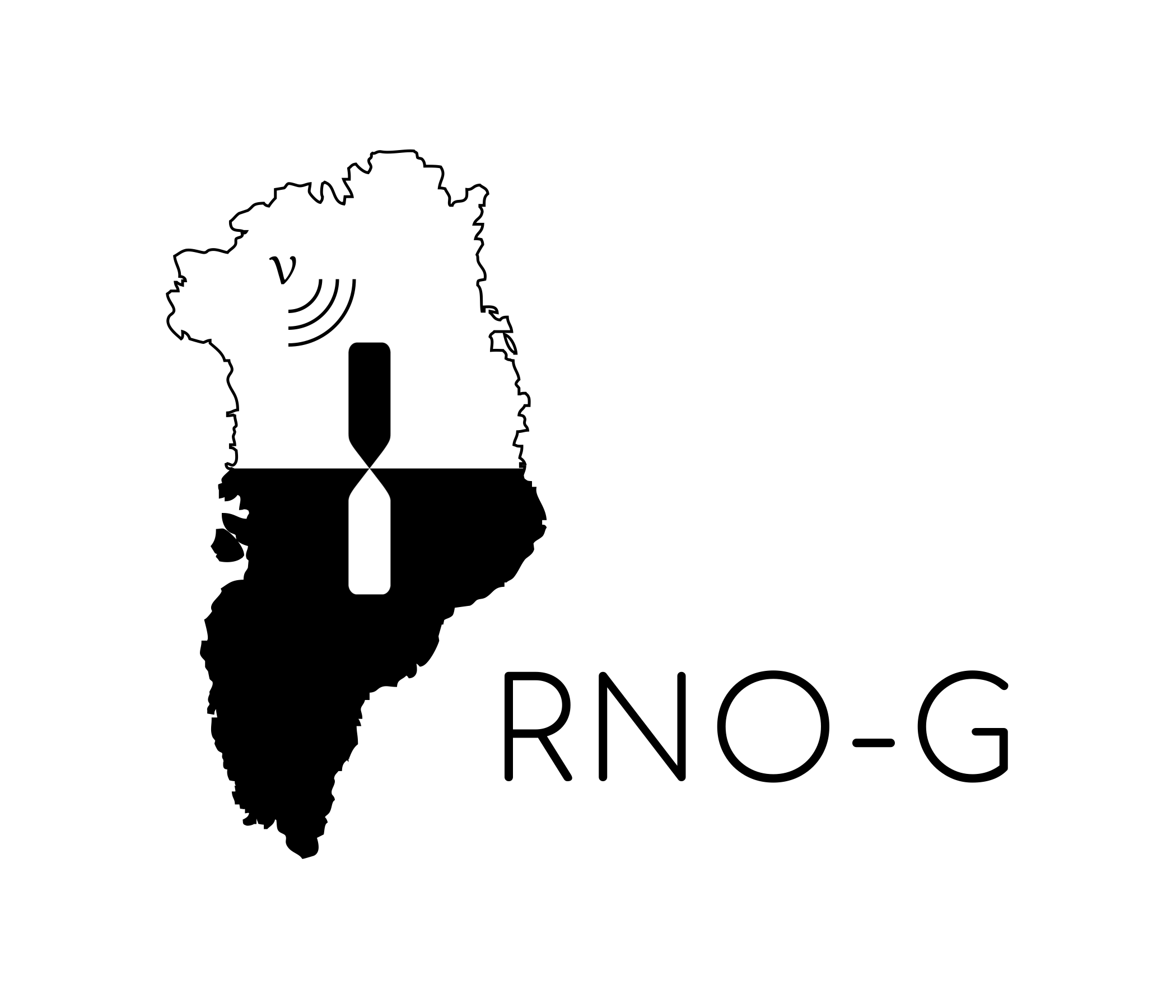}\\[6pt]
RNO-G Collaboration}

\author[1]{S.~Agarwal}
\author[2]{J.~A.~Aguilar}
\author[3]{N.~Alden}
\author[1]{S.~Ali}
\author[4]{P.~Allison}
\author[5,6]{M.~Betts}
\author[1]{D.~Besson}
\author[7]{A.~Bishop}
\author[8]{O.~Botner}
\author[9]{S.~Bouma}
\author[10,11]{S.~Buitink}
\author[2]{R.~Camphyn}
\author[7]{J.~Chan}
\author[2]{S.~Chiche}
\author[12]{B.~A.~Clark}
\author[1]{K.~Couberly}
\author[1]{D.~Dakroub}
\author[13]{K.~D.~de Vries}
\author[3]{C.~Deaconu}
\author[14]{P.~Giri}
\author[8,15]{C.~Glaser}
\author[4]{H.~Gui}
\author[8]{A.~Hallgren}
\author[16]{J.~C.~Hanson}
\author[14]{S.~Hassiki}
\author[17]{K.~Helbing}
\author[5,18]{B.~Hendricks}
\author[19,9]{J.~Henrichs}
\author[8]{N.~Heyer}
\author[1]{C.~Hornhuber}
\author[19]{E.~Huesca Santiago}
\author[4]{K.~Hughes}
\author[19,9]{A.~Jaitly}
\author[7]{A.~Karle}
\author[7]{J.~L.~Kelley}
\author[9]{C.~Kopper}
\author[2,13]{M.~Korntheuer}
\author[19,20]{M.~Kowalski}
\author[14]{I.~Kravchenko}
\author[5,18]{R.~Krebs}
\author[7]{M.~Kugelmeier}
\author[5,18]{D.~Kullgren}
\author[9]{R.~Lahmann}
\author[14]{C.-H. Liu}
\author[4]{Y.~Liu}
\author[21]{M.~J.~Marsee}
\author[11]{K.~Mulrey}
\author[7]{M.~Muzio}
\author[19,9]{A.~Nelles}
\author[22]{A.~Novikov}
\author[4]{A.~Nozdrina}
\author[3]{E.~Oberla}
\author[22]{N.~Punsuebsay}
\author[8]{M.~Ravn}
\author[4]{Z.~Riesen}
\author[17]{A.~Rifaie}
\author[23]{D.~Ryckbosch}
\author[2]{F.~Schl{\"u}ter}
\author[13,24]{O.~Scholten}
\author[9]{P.~Schriefer}
\author[22]{D.~Seckel}
\author[1]{M.~F.~H.~Seikh}
\author[19,9]{Z.~S.~Selcuk}
\author[23]{J.~Stachurska}
\author[13]{J.~Stoffels}
\author[2]{S.~Toscano}
\author[7]{D.~Tosi}
\author[6]{J.~Tutt}
\author[13]{N.~van Eijndhoven}
\author[3]{A.~G.~Vieregg}
\author[12]{A.~Vijai}
\author[19,9]{H.~Warnhofer}
\author[5,6]{D.~Washington}
\author[5,6,18]{C.~Welling}
\author[21]{D.~R.~Williams}
\author[3,19]{P.~Windischhofer}
\author[5,6,18]{S.~Wissel}
\author[9]{A.~Zink}
\affiliation[1]{University of Kansas, Dept.\ of Physics and Astronomy, Lawrence, KS 66045, USA}
\affiliation[2]{Universit\'e Libre de Bruxelles, Science Faculty CP230, B-1050 Brussels, Belgium}
\affiliation[3]{Dept.\ of Physics, Dept.\ of Astronomy \& Astrophysics, Enrico Fermi Inst., Kavli Inst.\ for Cosmological Physics, University of Chicago, Chicago, IL 60637, USA}
\affiliation[4]{Dept.\ of Physics, Center for Cosmology and AstroParticle Physics, Ohio State University, Columbus, OH 43210, USA}
\affiliation[5]{Dept.\ of Physics, Pennsylvania State University, University Park, PA 16802, USA}
\affiliation[6]{Dept.\ of Astronomy and Astrophysics, Pennsylvania State University, University Park, PA 16802, USA}
\affiliation[7]{Wisconsin IceCube Particle Astrophysics Center (WIPAC) and Dept.\ of Physics, University of Wisconsin-Madison, Madison, WI 53703,  USA}
\affiliation[8]{Uppsala University, Dept.\ of Physics and Astronomy, Uppsala, SE-752 37, Sweden}
\affiliation[9]{Erlangen Centre for Astroparticle Physics (ECAP), Friedrich-Alexander-University Erlangen-N\"urnberg, 91058 Erlangen, Germany}
\affiliation[10]{Vrije Universiteit Brussel, Astrophysical Institute, Pleinlaan 2, 1050 Brussels, Belgium}
\affiliation[11]{Dept.\ of Astrophysics/IMAPP, Radboud University, PO Box 9010, 6500 GL, The Netherlands}
\affiliation[12]{Dept.\ of Physics, University of Maryland, College Park, MD 20742, USA}
\affiliation[13]{Vrije Universiteit Brussel, Dienst ELEM, B-1050 Brussels, Belgium}
\affiliation[14]{Dept.\ of Physics and Astronomy, Univ.\ of Nebraska-Lincoln, NE, 68588, USA}
\affiliation[15]{Dept.\ of Physics, TU Dortmund University, Dortmund, Germany}
\affiliation[16]{Whittier College, Whittier, CA 90602, USA}
\affiliation[17]{Dept. of Physics, University of Wuppertal  D-42119 Wuppertal, Germany}
\affiliation[18]{Center for Multimessenger Astrophysics, Inst.\ of Gravitation and the Cosmos, Pennsylvania State University, University Park, PA 16802, USA}
\affiliation[19]{Deutsches Elektronen-Synchrotron DESY, Platanenallee 6, 15738 Zeuthen, Germany}
\affiliation[20]{Institut f\"ur Physik, Humboldt-Universit\"at zu Berlin, 12489 Berlin, Germany}
\affiliation[21]{Dept.\ of Physics and Astronomy, University of Alabama, Tuscaloosa, AL 35487, USA}
\affiliation[22]{Dept.\ of Physics and Astronomy, University of Delaware, Newark, DE 19716, USA}
\affiliation[23]{Ghent University, Dept.\ of Physics and Astronomy, B-9000 Gent, Belgium}
\affiliation[24]{Kapteyn Institute, University of Groningen, PO Box 800, 9700 AV, The Netherlands}

\emailAdd{ryan.j.krebs@psu.edu, authors@rno-g.org}

\abstract{The Radio Neutrino Observatory in Greenland (RNO-G) is a neutrino detector under construction at Summit Station, with 8 out of a planned 35 stations currently deployed. We have designed and deployed a new phased array (PA) trigger based on delay-and-sum beamforming and power integration. This trigger improves detector performance by suppressing thermal noise and better targeting neutrino-induced Askaryan signals. The new PA trigger has been deployed since the 2025 season. The trigger performance has been characterized using test pulses and calibration pulsers both in the lab and in-situ, and we find across these tests a 25\% average reduction in the signal-to-noise ratio (SNR) needed to trigger on signals. Simulations are shown to be representative of the detector, and simulated trigger efficiencies are within 10\% of measured data. Following the in-situ trigger validation, we use data-driven trigger performance to inform simulations of the detector effective volumes. The PA trigger increases our effective volume significantly, by over a factor of 2 below 0.1\,EeV and at least a factor of 1.3 at the highest energies of 100\,EeV.}

\keywords{Large detector systems for particle and astroparticle physics; Neutrino detectors; Antennas; Interferometry; Performance of High Energy Physics Detectors}

\arxivnumber{1234.56789} % Only if you have one

\begin{document}
\maketitle
\flushbottom

\section{Introduction}

% intro to rno-g and neutrino detection
The Radio Neutrino Observatory in Greenland (RNO-G) is a hybrid radio experiment combining the designs of shallow arrays \cite{Barwick:2014boa, Nelles:2015tch, RICE:2001ayk, Kravchenko:2012ni} and in-ice arrays \cite{Allison:2011wk, ARA:2015wxq, ARA:2019wcf, Allison:2018ynt} for the detection of ultra-high energy (UHE) neutrinos ($\geq$100 PeV). Located at the summit of the Greenlandic ice sheet, the stations sit atop $\sim$3\,km of ice. RNO-G is planned for 35 independent stations with 8 stations currently deployed. The stations are self-triggered on radio frequency (RF) signals using thermal noise riding triggers. Neutrino events are expected to be quite rare and their radio signature, produced by the Askaryan effect \cite{Askaryan:1961pfb,Askaryan:1965}, will be comparable to the thermal noise levels.  Thus, implementing a lower threshold trigger is critical to the success of RNO-G. This is particularly true near the energy threshold of the experiment ($\sim10^{16}$\,eV) where the expected flux is higher, but the emitted radio signal is weaker.

% rno-g stations
Each station is built with a centrally located data acquisition (DAQ) box, 3 boreholes instrumented with antennas, and shallow antennas embedded just under the ice. The first seven stations (11, 12, 13, 21, 22, 23, 24) are designed with 7 vertically polarized (Vpol) and 2 horizontally polarized (Hpol) antennas on the main ``power'' string, 2 Vpol antennas and 1 Hpol antenna on each ``helper'' string, and a shallow array containing 3 log-periodic dipole antennas (LPDA) pointing upwards and 6 LPDAs facing downwards. A calibration pulser is deployed on each helper string deep in the ice and a shallow pulser located just under the surface. The most recent station, 14, begins a transition to a slightly adjusted antenna layout with 8 Vpol antennas on the power string, adding in a shallower Vpol, and converting the shallow array to 4 upwards facing LPDAs and 4 downwards facing LPDAs.

The deepest 4 Vpol antennas on the power string are called the phased array (PA). Each in-ice antenna is connected to a front-end amplifier at the antenna, In-Ice Gain and Low Power Unit (IGLU), that connects to a second-stage amplifier, Downhole Receiver and Amplifier Board (DRAB), in the DAQ box. Cables between the IGLU and DRAB are single mode Radio Frequency over Fiber (RFoF) cables. The shallow channels are simpler and only contain one set of amplifiers per channel with LMR-400 coaxial cables connecting the antennas to the amplifiers.

The DAQ contains complementary digitizer boards: the RADIANT for recording of the full bandwidth signals and auxiliary triggers on the shallow channels, and the FLOWER for limited bandwidth digitization and triggering of the PA channels. Each board interfaces with the single board computer (SBC) to control the electronics, read out data, and transfer data off station to a data storage server. A complete description of the detector can be found in Refs.~\cite{RNO-G:2020rmc,RNO-G:2025inst}. RNO-G stations use various triggers to capture data. The RADIANT triggers operate on the shallow LPDAs and primarily trigger on above-surface signals. The FLOWER triggers are primarily used to capture neutrino signals using the PA antennas. Software triggers, initiated by the SBC, are used to characterize the noise background. % with many supplementary updates in \cite{RNO-G:2021upe, RNO-G:2023jlv, RNO-G:2023otj,  RNO-G:2023fkv, RNO-G:2023xyg, RNO-G:2023jct, RNO-G:2023snh, Cataldo:2023qbe, RNO-G:2023fkv, Windischhofer:2024xgw, Krebs:2024Sy, Agarwal:2024cl}.

% the phased array and trigger hardware
The PA antennas are nominally deployed between 96-98\,m below the surface and are positioned with 1\,m separation between the centers of each antenna. These channels are split and fed to the RADIANT and the FLOWER. The FLOWER contains two HMCAD 1511 streaming digitizers with 8-bit samples at a sampling rate of 472\,MHz. Samples are streamed to a low-power Cyclone V Field Programmable Gate Array (FPGA). This configuration allows for the use of a digital PA system. The trigger bandwidth has been limited to up to 236\,MHz, having been low-pass filtered from the full bandwidth of 90-800\,MHz seen at the RADIANT. This reduced bandwidth benefits the relative neutrino signal strength to the background thermal noise~\cite{Glaser_2021_triggers}. Channels are gain adjusted so the root mean square (RMS) voltage of thermal noise in the individual channels is at least 5 ADU. 

% the hi-lo trigger
The trigger is an action that tells the RADIANT to read out a 853.3\,ns snapshot of all channels. The trigger running for the 2021-2024 seasons was a voltage-threshold type. It detects signals above and below a threshold voltage (hi-lo) in a $\sim$12.7\,ns window (6 samples) on a single channel. Then coincidences between channels are checked in a window of $\sim$42.4\,ns (20 samples) with 2 of 4 channels over-threshold constituting a trigger. This trigger's measured 50\% efficiency point of 4$\sigma$~\cite{RNO-G:2025inst}, where $\sigma$ is the root mean square voltage of thermal noise, falls short of the assumed 2$\sigma$ trigger in Ref.~\cite{RNO-G:2020rmc}. This drives the need for an improved trigger that can detect weak neutrino-like signals on a thermal noise background.%Therefore, we need to adopt a better trigger to suppress thermal noise and target neutrino-like signals.

% beamforming
Beamforming is the process of adding waveforms between similar channels so that real signals from a particular direction add coherently, but random thermal noise adds incoherently. This is one of the basic concepts of radio astronomy. However, it is relatively new to the radio neutrino community and is currently being adopted by more radio detector experiments. It has been shown that beamforming and performing the trigger logic on the beamformed traces can significantly improve the sensitivities of radio neutrino detectors \cite{Vieregg:2015baa, Avva:2016ggs, Allison:2018ynt, Southall:2022yil, Zeolla:2023khf, PUEO:2020bnn}. This work presents beamforming as the adopted strategy to improve detector performance. Due to the impulsive nature of Askaryan radio emission, we use time-based beamforming, as was done previously in Ref.~\cite{ARA:2019wcf}. This means that for each beam the channels are delayed by some amount and added together to create a ``coherent'' sum, or the beamformed traces. This helps to suppress thermal noise compared to real signals arriving at the beam location. For ideal beamforming the signal to noise ratio (SNR) of the coherent sum is a factor of $\sqrt{n}$ larger than a single channel's SNR, where $n$ is the number of channels in the sum. 

\begin{figure}[htbp]
    \centering
    \includegraphics[width=1\textwidth]{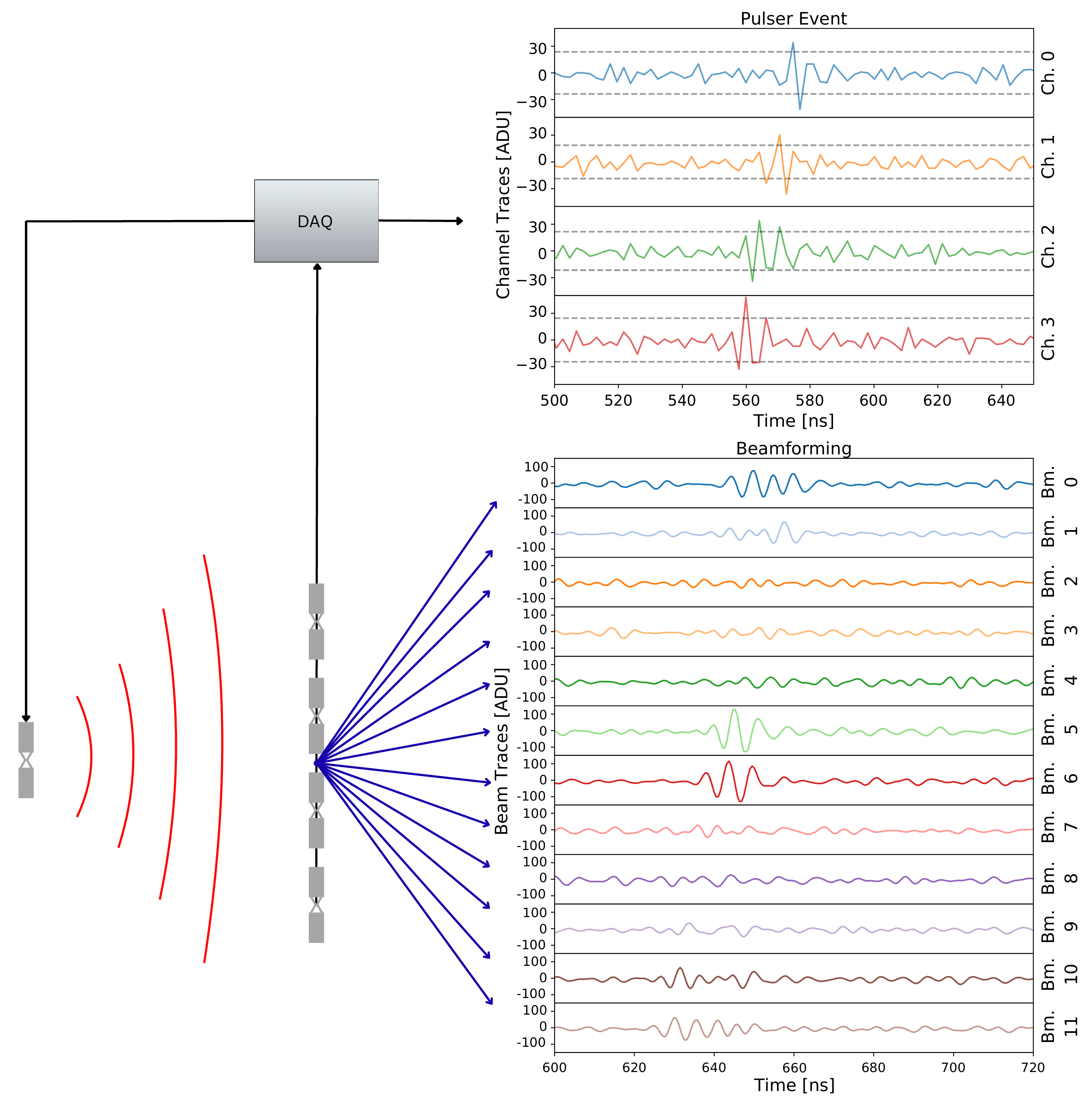}
    \caption{The main testing method for trigger performance is through in-situ pulsers. This shows an example pulser event recorded in the FLOWER and each individual beam trace after upsampling and beamforming is performed. The beam direction aligning nearest to the received direction of the RF signal has the largest coherence and shows the largest signals. The most vertical beams (beams 0, 1, 10, 11) show distinct pulses from each channel with no interference. The center beams (beams 5, 6) show the maximum signal where the signals coherently align. The intermediary beams (beams 2, 3, 4, 7, 8, 9) show limited signals where they interfere destructively.}
    \label{fig:diagram}
\end{figure}

To illustrate the concept, Figure~\ref{fig:diagram} shows an example of beamforming to the on-station calibration pulsers. Digitized waveforms are shown with example hi-lo trigger thresholds on each channel. Beamforming prioritizes specific incident directions by choosing the beam delays that align signals in time using the expected propagation time through the ice and cables. Beamformed traces are shown and it can be seen that the beams pointing back to the pulsers are the strongest.

In addition, real antennas and signal chains are dispersive which extends the nanosecond-scale bipolar electric field pulses of Askaryan emission in time. Ref.~\cite{Glaser_2021_triggers} explores different triggering methods to neutrino signals through signal chains and antennas of different levels of dispersion. Voltage threshold triggers, which were used on RNO-G previously, perform better to highly impulsive, short-duration, signals, whereas power integration triggers perform better on longer signals. Various triggering methods are explored in this work for RNO-G.

RNO-G was designed with beamforming in mind, having adapted the ARA PA~\cite{Allison:2018ynt} into the RNO-G station layout. In Ref.~\cite{RNO-G:2025inst} we presented a prototype beamformed trigger that, as measured, did not perform better than the hi-lo trigger, although simulations suggested it would do better with neutrino signals than the calibration pulser. Between then and the start of the 2025 season, various improvements were made to the firmware and station operation to increase the performance of this trigger.

% outline paper
The work presented here details the PA trigger implementation, validates both triggers' performance using both lab and on-station testing, and estimates the detector's performance increase to the previous hi-lo trigger using the ratio of the single-station effective volumes. In Section~\ref{sec:simulations} we will introduce the simulations needed to model neutrino signals, compare the calibration pulser measurements to a simulated version, and to estimate the performance gain with the new trigger. Section~\ref{sec:design} details the specific trigger implementation that was designed and run on the detector since the 2025 season and compares that to various other triggering methods. Section~\ref{sec:validation} describes the data collection methods, both in the lab and on station, and shows experimental results for validating the hi-lo and PA trigger in the lab using test pulses and on-station with the embedded calibration pulsers. Lastly, Section~\ref{sec:effective_volume} uses the extracted trigger performance to help estimate the detector-level performance of the single-station effective volume ratios.

\section{Simulation setup}
\label{sec:simulations}

To generate comparisons for the measurements presented further in this work, we use the simulation framework, \texttt{NuRadioMC} (and its sub-package NuRadioReco)~\cite{Glaser:2019cws,Glaser_2019}. \texttt{NuRadioMC} is used to generate neutrino events or test calibration signals, propagate signals through ice, and model the RNO-G detector response and trigger. This section shows Askaryan pulses and compares them to the on-station calibration pulsers, justifying their use as a proxy to estimate the trigger's performance to neutrinos. We then recreate the validation tests using on-station pulsers for comparison with measurements. Further usage of \texttt{NuRadioMC} is detailed in Section~\ref{sec:effective_volume}.

\subsection{Simulated radio emission} 
\label{sec:nu_sims}

Neutrino interactions in polar ice are expected to generate Askaryan radiation, resulting from the coherent radio emission of net negative charge build up in a compact particle shower. In this work, we compare the performance of the trigger to the ARZ2020 model~\cite{ARZ2020} of Askaryan emission. It is a semi-analytic treatment which reproduces the time-domain electric fields from microscopic computation modeling of Askaryan radiation~\cite{PhysRevD.45.362} by convolving the charge excess profile with a parameterized emission profile. The ARZ2020 model is the most up-to-date model of Askaryan emission, and includes the details needed to model both hadronic and electromagnetic particle showers from neutrino interactions.

RNO-G is sensitive to neutrinos across a large volume of ice, and radio propagation effects are therefore imprinted on the received radio signature. The complex dielectric constant and loss tangent of the ice can alter the signal shape and attenuate the signal amplitude. 
We use the most accurate attenuation length model currently available, Greenland-3 (GL3) in \texttt{NuRadioMC}, derived from direct attenuation length measurements~\cite{GL3}. The model used for the index of refraction of the ice, which governs ray bending, is a single-exponential model derived from data in Ref.~\cite{cosmin_ice}. 

\begin{figure}
    \centering
    \includegraphics[width=.8\textwidth]{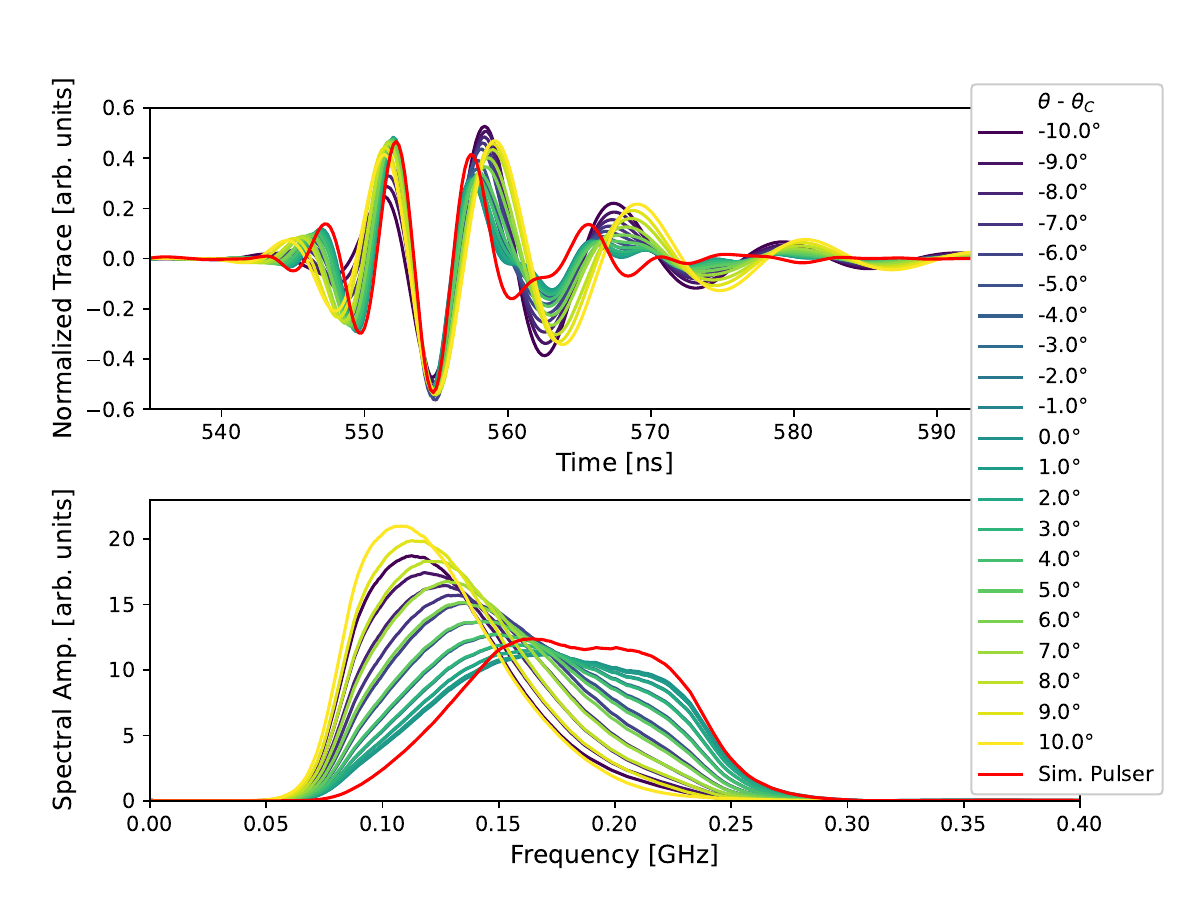}
    \caption{Askaryan pulses generated with the ARZ2020 \cite{ARZ2020} model propagated through 1\,km of ice \cite{GL3} and received at a Vpol antenna on boresight. Various view angles from the Cherenkov angle are shown to compare the relative frequency composition. Top: Normalized traces compared with the simulated pulser model. Normalization is performed on the time-domain traces by dividing the full trace by the peak-to-peak voltage. Bottom: Spectrum of the normalized traces compared with a simulated pulser.}
    \label{fig:nu_traces}
\end{figure}

The neutrino signals vary depending on the angle of emission away from the Cherenkov angle, defined here as the view angle. The signals have the flattest spectrum directly on the Cherenkov angle ($\theta = 0^{\circ}$), losing high frequencies when viewing further off-cone. Figure~\ref{fig:nu_traces} shows amplitude normalized traces on the trigger board before digitization using an example hadronic shower from the semi-analytic emission model ARZ2020~\cite{ARZ2020}. We note that Askaryan pulse amplitudes are strongest on-cone and decrease further away from the Cherenkov angle. We expect off-cone geometries to encompass more of the event population space. We also compare these signals to a simulation of the in-situ calibration pulsers. 

The simulated model of the calibration pulser is generated by using the measured voltage output of the calibration pulser board. This measured pulse is sent through an emitting Vpol antenna that is modeled with the finite-difference time-domain electromagnetic simulation software XFDTD~\cite{XF}. The signals are propagated to the receiving antennas and pass through a measured model of the signal chains before being digitized. An example of the received signal at the DAQ can be seen in Figure~\ref{fig:nu_traces} where the test calibration signal is compared to expected neutrino Askaryan pulses. While the calibration pulsers have more power at high frequencies than on-cone neutrino signals, they are quite similar in both time and frequency; therefore the trigger performance to expected neutrino signals can be estimated with the in-ice calibration pulses.

\subsection{Simulated trigger and setup} 
\label{sec:pulser_sims}

The detector simulations use realistic behavior of the FLOWER's digitization and trigger arithmetic that is performed on an FPGA. We note that the low-pass filter on the FLOWER is insufficient to filter the frequency band to below the Nyquist frequency, and in this case, the aliasing effects are included in the simulations of the FLOWER digitization. In an effort to overhaul the detector simulations, the trigger code was updated to include the characteristics of firmware: fixed-point integer math, data bit depths, and data bit depth saturation. The triggers use an adjustable threshold. These thresholds correlate to the detector performance where lower thresholds are able to trigger on lower amplitude signals, but they also affect the overall event and data rates. To match typical event rates on station (due to limitations in readout rates and data transmission budgets), the simulation-derived thresholds are set such that the station event trigger rates on simulated thermal noise traces are approximately 1\,Hz. 

To complement the on-station trigger testing in Section~\ref{sec:on-station} we simulate trigger efficiencies across a range of elevation angles to the PA. The trigger efficiency is the probability that a specific signal triggers the detector, which is not deterministic due to noise. We use a similar setup to the real setup in Section~\ref{sec:validation} by placing an emitting antenna on a vertical line 35\,m away from the PA. The positions are then scanned along the vertical line. We verified that the simulated pulser signal matches with the measured pulse shapes. Pulse amplitudes are randomly generated to sufficiently span SNRs from near zero up to saturating the FLOWER's digitizer.

\section{Trigger design}
\label{sec:design}
 
 The triggers presented here all perform beamforming, which is a well-established method to enhance the SNR of weak signals, thereby lowering the trigger and energy threshold of a neutrino experiment. The implemented PA trigger presented here is composed of 3 main algorithmic steps: (1) digital upsampling, where the 472\,MHz waveforms are upsampled to 1888\,MHz using a Finite Impulse Response (FIR) filter; (2) beamforming using predefined channel delays (delay and sum); and (3) the calculation and comparison of an integrated power to a threshold. Each step is discussed in full, along with how various design options affect the trigger performance and resource usage.

%A general structure for this style of signal agnostic digital triggers includes a pre-beamforming filter, beamforming, post-beamforming filter, trigger arithmetic, and the threshold comparison. The pre-beamforming filtering can be anything from digital filtering, de-dispersion, or up/down sampling. The beamformer generates $n-$beamformed traces from the input channels using delay and summing. The post-beamforming filter can be any style of filters that further improves signal detection. Then the trigger arithmetic to determine over-threshold signals can be simple voltage threshold, hi-lo, power integration, cross correlation, Hilbert enveloping, or others. Finally, the output is compared to noise riding thresholds to determine if a trigger happens.

The behavior of the register transfer level (RTL) description of the implemented firmware is compared to the detector simulation model to verify correct functionality. RTL is the low-level structure of the firmware which describes the physical circuits being built on the FPGA. RTL simulations are performed with the open-source VHDL simulator GHDL~\cite{GHDL}.

\subsection{Upsampling}
\label{sec:upsampling}
Samples from the ADCs are first sent through an initial upsampling stage. Due to the antenna spacing and sampling rate, we need to perform upsampling to better recover signal peaks in the traces and to better pack beams into the chosen angular region. In a system sampling just at the Nyquist frequency, the transition from constructive to destructive interference is as small as a single sample. Errors from nearest-neighbor integer sample delays are reduced with a higher sampling rate. Generating more tightly spaced beams guarantees all incident angles of interest are in-phase in at least one beam. 

We employ digital upsampling using a FIR method. The algorithm works by zero-stuffing the original trace with $n-1$ zeros (for upsampling at a factor of $n$) between each physical sample, then the zero-stuffed signal is sent through a FIR low-pass filter to remove the spectral images above the original Nyquist frequency. Upsampling by a factor of 4 was found to be a good trade-off between resource usage and detector performance, outperforming upsampling by a factor of 2.

\begin{figure}
    \centering
    \includegraphics [width=1\textwidth]{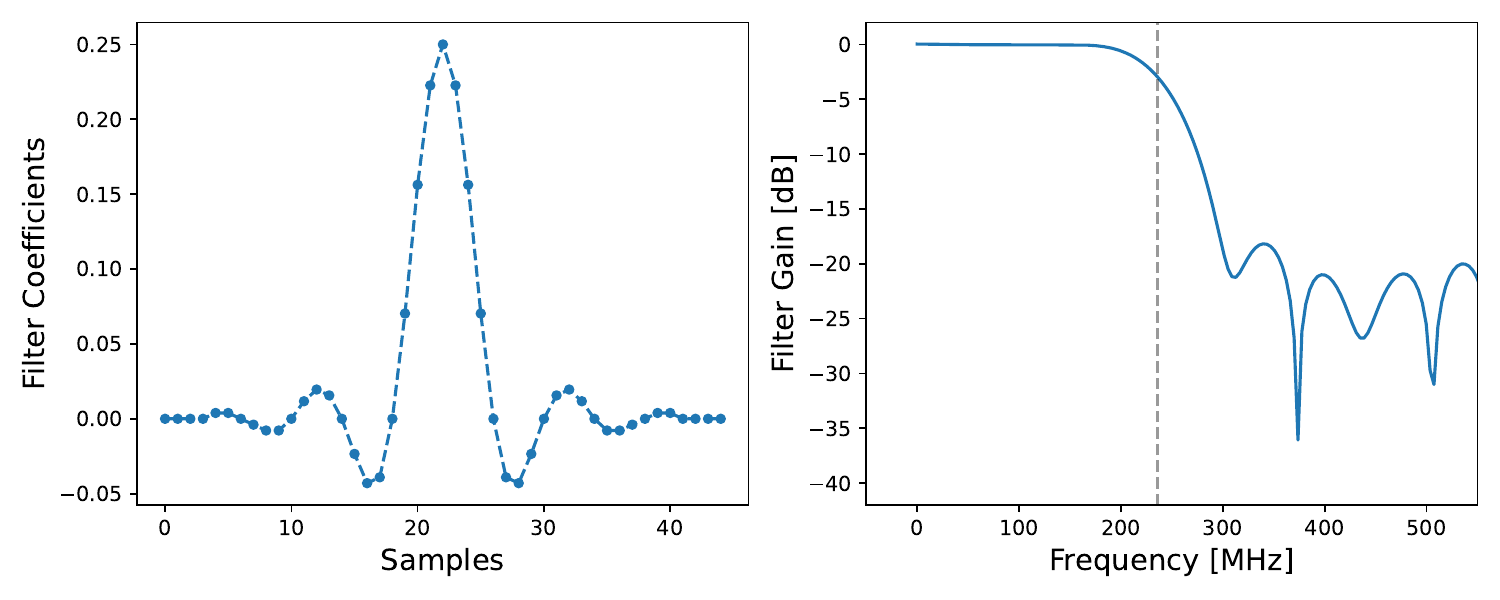}
    \caption{When digitally upsampling, the zero-stuffed waveform is passed through a finite impulse response (FIR) low-pass filter to remove spectral images above the original Nyquist frequency. Left: Time domain coefficients of the low-pass filter. Right: Frequency response of the low-pass filter with a 3\,dB cutoff around 236 MHz (dashed line).}
    \label{fig:filter}
\end{figure}

When upsampling by a factor of 4, the low-pass filter is chosen to be a quarter-band Hamming filter with 45 coefficients at 8 bits each. The filter coefficients are quantized to integer values by rescaling and rounding the floating point coefficients. Fixed-point math (integer math) is necessary to reduce the resource cost. The bit length of the coefficients is kept big enough to have good representation, but small enough so that many small coefficients are forced to 0, further saving resources. Further optimizations to force coefficients to sums of factors of 2 are possible, as bit-shifting samples and adding small multiples of the samples is much more efficient than doing multiplies. It is not implemented in the current state of the design, but further optimizations to this end are possible. Figure~\ref{fig:filter} shows the chosen filter coefficients and the frequency response of the filter. Increased bit depths and more coefficients can produce sharper cutoffs to keep out-of-band spectral images limited, but reconstructing a perfect representation of the signal is unneeded for triggering.

\begin{figure}
    \centering
    \includegraphics [width=.8\textwidth]{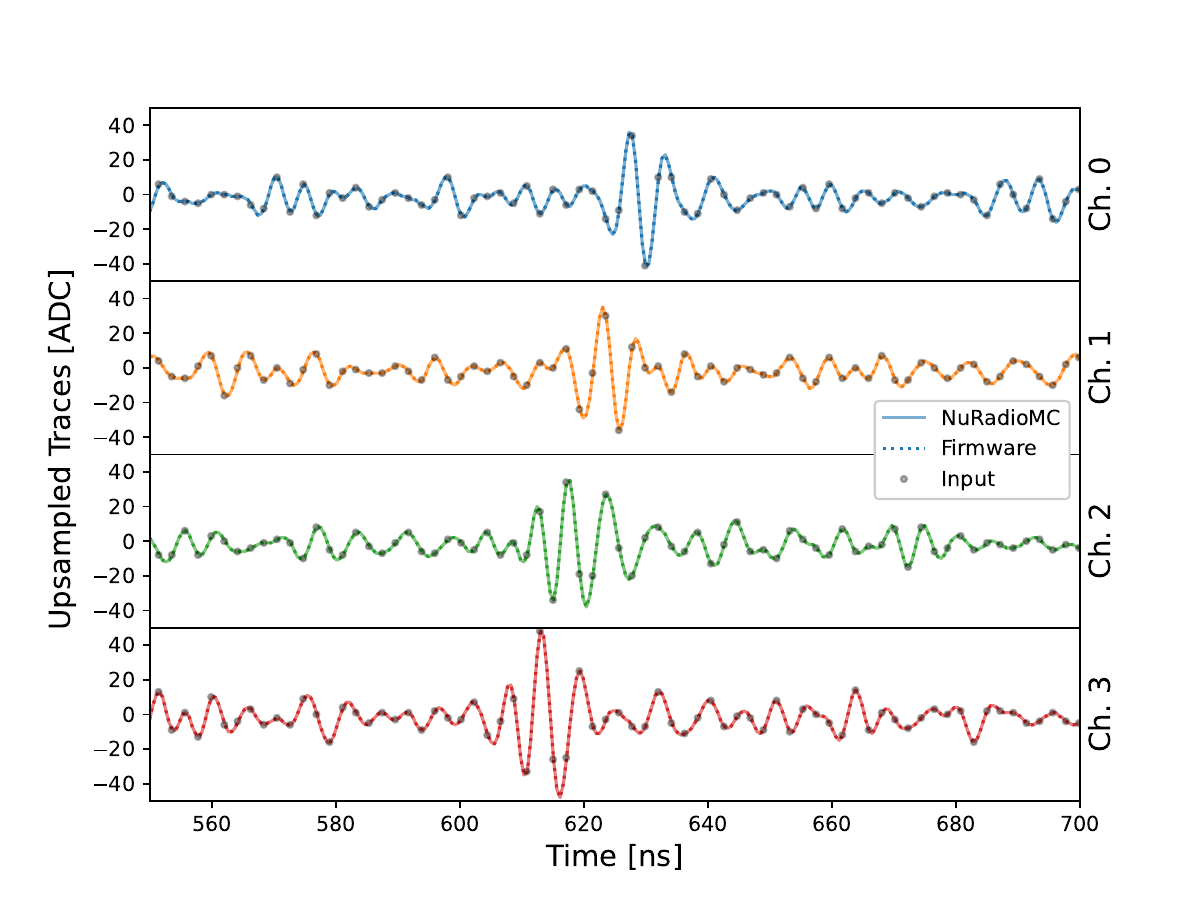}
    \caption{Example pulser event showing the upsampling as performed in the \texttt{NuRadioMC} detector simulations and in the behavioral model of the firmware.}
    \label{fig:upsampling}
\end{figure}

Figure~\ref{fig:upsampling} shows an example of upsampling a pulser signal performed in both the RTL simulations and \texttt{NuRadioMC}, which agree. 

On the FPGA, where calculations are performed on new samples being streamed in on each clock cycle, the data on this step increases the 4 samples $\times$ 4 channels to 16 samples $\times$ 4 channels. Input and output samples of this stage are constrained to 8-bit signed integers only. Saturation of the input digitized samples for very large signals can affect upsampling behavior, although such large signals should trigger regardless.

We tested an intermediary step using either de-dispersion (group delay equalization) or matched filtering before upsampling. De-dispersion, used previously on the PUEO trigger~\cite{PUEO:2020bnn}, removes the phase delays of the antenna and signal chains so that dispersive signals become more impulsive. Matched filtering targets specific signal shapes using a convolution. Performance did not improve over the current implementation. The small dispersion of the Vpol antenna and signal chains does not make these techniques critical. While not implemented here, similar techniques targeting neutrino signals and background rejection may be revisited with more resourceful, but lower power, FPGAs in the future.

\subsection{Beamforming}
The output of the upsampling module is sent to the beamforming module. Beamforming is performed using a delay-and-sum method. For simplicity, we assume a plane wave approximation to calculate arrival times at a channel using Eq.~\eqref{eq:beam_locs}:

\begin{equation}
\label{eq:beam_locs}
    t_{\mathrm{beam}}(i)= \Big(\frac{n\cdot (d_{3}-d_{i})\cdot \sin (\theta_{\mathrm{beam}})}{c}-t_{\mathrm{cable}}(i)\Big)
\end{equation}

The arrival time at each channel, $t_{\mathrm{beam}}(i)$, for each beam is determined by calculating the time due to a plane wave arriving at each antenna subtracted by the cable delay time, $t_{\mathrm{cable}}(i)$. The indices $i$ are used to denote the channel. The propagation time delay depends on the incident angle, $\theta$, the depths of the antennas, $d$, the index of refraction at the depth of the PA antennas, n$\approx$1.75, and the speed of light in a vacuum, $c$. The index of refraction varies greatly at shallow depths~\cite{ice_index_2025}, but becomes uniform near and below the PA antenna depths~\cite{deep_ice_index_2024} allowing for the use of Eq.~\eqref{eq:beam_locs}.

The channel time delays are calculated with Eq.~\eqref{eq:beam_locs} using 12 beams located in equally spaced intervals in cos($\theta$) from -60\degree~to 60\degree~for the upsampled rate of 1888\,MHz. Beams are located at $\pm$60.00\degree, $\pm$45.12\degree, $\pm$33.44\degree, $\pm$23.18\degree, $\pm$13.66\degree, $\pm$4.52\degree. These angles cover the bulk of the antenna gain pattern found in Ref.~\cite{RNO-G:2025inst}. The arrival times for each beam and channel from Eq.~\eqref{eq:beam_locs} in units of time are manipulated to become a look back time in units of integer samples, $k_{\mathrm{beam}}(i)$, for use on the FPGA, given by Eq.~\eqref{eq:lookbacks}, where $f$ is the upsampled sampling rate of 1888\,MHz. The minimum arrival time for the set of channels is subtracted from each arrival time for a beam to reduce latency through this step.

\begin{equation}
    \label{eq:lookbacks}
    k_{\mathrm{beam}}(i) = \mathrm{int}\Big(f\cdot (t_{\mathrm{beam}}(i)-\mathrm{min}(t_{\mathrm{beam}}))\Big )
\end{equation}

While these delays can differ between stations, each antenna is carefully deployed with a relative variance of PA antenna depths of $\pm$ 3\,cm. Cable delays can differ as well, but are at most 400\,ps which is unexpected to change beamforming delays significantly. This allows us to use the same set of delays and firmware for all stations. The pipeline to produce station-specific firmware is in place in case future stations' hardware or geometry differ.

The beamformed traces, $V_{\mathrm{beam}(s)}$, are computed with Eq.~\eqref{eq:beam_sum} using each channel's trace, $V_{\mathrm{ch}}(s)$, and the look back time calculated previously. 

\begin{equation}
    \label{eq:beam_sum}
    V_{\mathrm{beam}}(s)=\sum_{\mathrm{i}=0,1,2,3}V_{\mathrm{i}}(s-k_{\mathrm{beam}}(i))
\end{equation}

We tested beam steering with adjustable look back times that are set by software and allow for small adjustments of $\ge$3 samples on top of a standard set. Real-time beam steering is not needed given a dense set of beams with sufficient coverage; it also uses 20\% more logic cells and an additional 1\,W of power. Due to the increased resource usage and the builtin beam prioritization using beam-specific thresholds, we choose to not implement this feature.

Sample delays span up to 72 samples per channel so a shift register is used to temporarily store channel samples. Then necessary samples are multiplexed and fanned out to form the summation needed for each beam. In this step, the 16 samples $\times$ 4 channels is increased to 16 samples $\times$ 12 beams. In the algorithm, the 8-bit signed integers naturally increase up to 10-bit signed integers and are then saturated down to 8-bit signed integers. Further saturation to lower bit lengths was considered to save resources, but keeping the original size of 8 bits was needed for highly impulsive signals on a background noise RMS voltage of 5\,ADU, discussed further in Section~\ref{sec:trigger_comparison}.

\subsection{Power integration}
The final step on the trigger is to envelope the beamformed traces and compare to a threshold. Simple thresholding without enveloping, digital Hilbert enveloping, and power integration enveloping were considered. We chose power integration as it targets the extended shapes of the off-cone neutrino signals to enhance the effective volume (see Section \ref{sec:trigger_comparison}). It has been used for other experiments as well \cite{ARA:2019wcf, Southall:2022yil} and is supported by Ref.~\cite{Glaser_2021_triggers}.

The power integration step first squares the samples of the beamformed traces by using a lookup table (LUT), embedded on Block Random Access Memory (BRAM). The LUT uses the 8-bit samples as an input and outputs the 16-bit powers. Full calculations of the power can be performed in logic slices or on digital signal processing (DSP) blocks instead, but BRAM resources are available. This reduces logic slice usage and power. Then the powers are summed inside of an integration window of $\sim$12.7\,ns (24 samples). The window slides across the waveforms with a step of $\sim$2.1\,ns (4 samples) to minimize missing peak integrated powers. The data bit depth of the power sum is allowed up to 24 bits to not overflow, although we should not expect overflows to occur in normal operations where signals are comparable to the static background noise level. We then divide the integrated power by 32 to calculate an ``average'' power --- this is performed by bit shifting down by 4 bits. The reduced integrated power is then compared to the trigger and servo thresholds. Thresholds are stored as 12-bit unsigned integers so that the trigger and servo thresholds pack efficiently into 24-bit registers.

\begin{figure}[htbp]
    \centering
    \includegraphics[width=.65\textwidth]{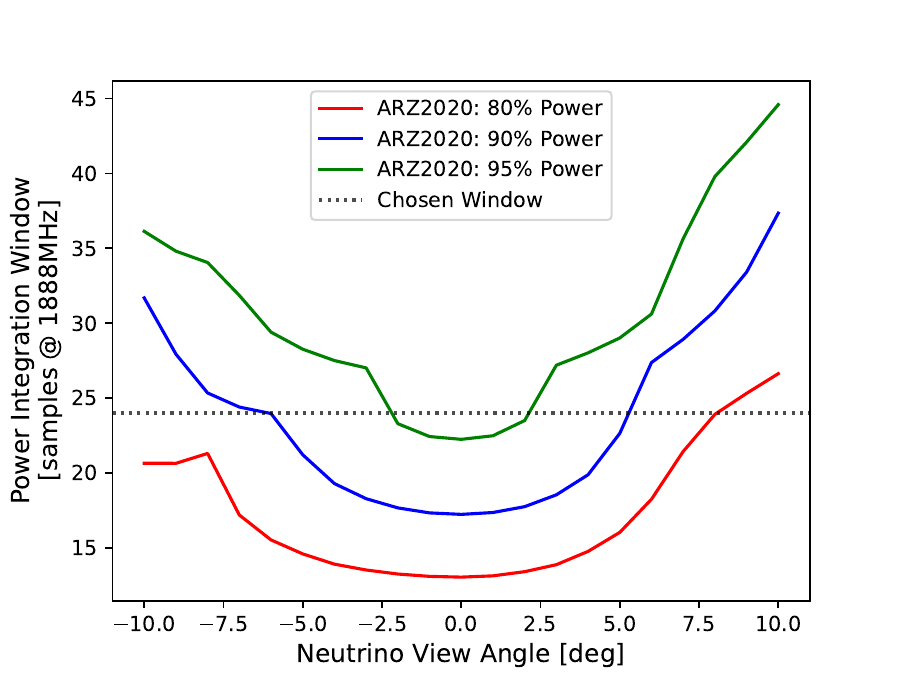}
    \caption{Power integration windows, in samples, needed to reach 80\%, 90\%, and 95\% of the full power of neutrino signals at varying view angles from the Cherenkov angle for the ARZ2020~\cite{ARZ2020} Askaryan models.}
    \label{fig:power_nu_eff}
\end{figure}

Various integration lengths were tested in detector simulations. Figure~\ref{fig:power_nu_eff} compares the percent power contained in different integration lengths of various view angles to the Cherenkov angle. The chosen integration length of $\sim$12.7\,ns (24 samples) contains up to 90\% power of the entire waveform. Section~\ref{sec:trigger_comparison} compares different power windows to station performance and finds negligible difference between windows.  

Table \ref{tab:resource_usage} compares the resource usage of the various steps and the full trigger with the Cyclone V FPGA (model 5CEFA5F23I7N). The PA trigger uses more resources than the simple hi-lo trigger. In the final post-fit optimization, physical placement of logic, the PA trigger reduces to $\sim$40\% of the FPGA's logic usage. The power used by the PA adds $\sim$1\,W on top of the base FLOWER power usage of $\sim$3\,W. We also chose to use logic-slice based math on the FIR filters that increases resource usage, but does reduce total power by $\sim$1\,W by not using the DSP blocks.

\begin{table}[htbp]
    \centering
        \caption{Resource usage broken down by trigger type. Trigger usage is pre-fit, pure RTL logic usage, before the design is resource optimized. Post-fit, after physical placement and condensing of logic cells, is after optimization which efficiently packs resources and reduces the overall resource usage.}
    \begin{tabular}{|c|c|c|c|}
         \multicolumn{4}{c}{Firmware Resource Usage}\\
    \hline
         Entity&  ALUTs (\%) &  Registers (\%)&  BRAM [bits]\\
    \hline
         Upsampling&  7981 (27.4)&  5664 (9.7)&  72\\
         Beamforming&  5184 (17.8)&  2920 (5.0)&  0\\
         Power Integration&  5616 (19.3)&  3804 (6.5)&  540k\\
         Pre-Fit PA Total&  20622 (70.9)&  13652 (23.5)&  540k\\
         Post-Fit PA Total& 17798 (61.2)& 34387 (59.1)& 833k\\
        \hline
         Hi-Lo Total& 669 (2.3)& 708 (1.2)& 0\\
         \hline
         Available& 29080& 58160& 4567k\\
    \hline
    \end{tabular}

    \label{tab:resource_usage}
\end{table}

\subsection{Trigger comparison}
\label{sec:trigger_comparison}

Across the beamformed triggers we can tune the upsampling factor to better form beams across elevation angles and to recover the peaks in an undersampled signal. Higher upsampling factors and better-performing upsampling use more FPGA resources. We can then test the effects of different saturation during beamforming. Limiting the bit depth of samples can help to limit resource usage. With the power integration trigger we can change the integration window to target more dispersive signals. Ideal upsampling and beamforming, unrealizable on a limited FPGA are shown for comparison as these were the original assumptions in Ref.~\cite{RNO-G:2020rmc}.

\begin{figure}
    \centering
    \includegraphics[width=1\textwidth]{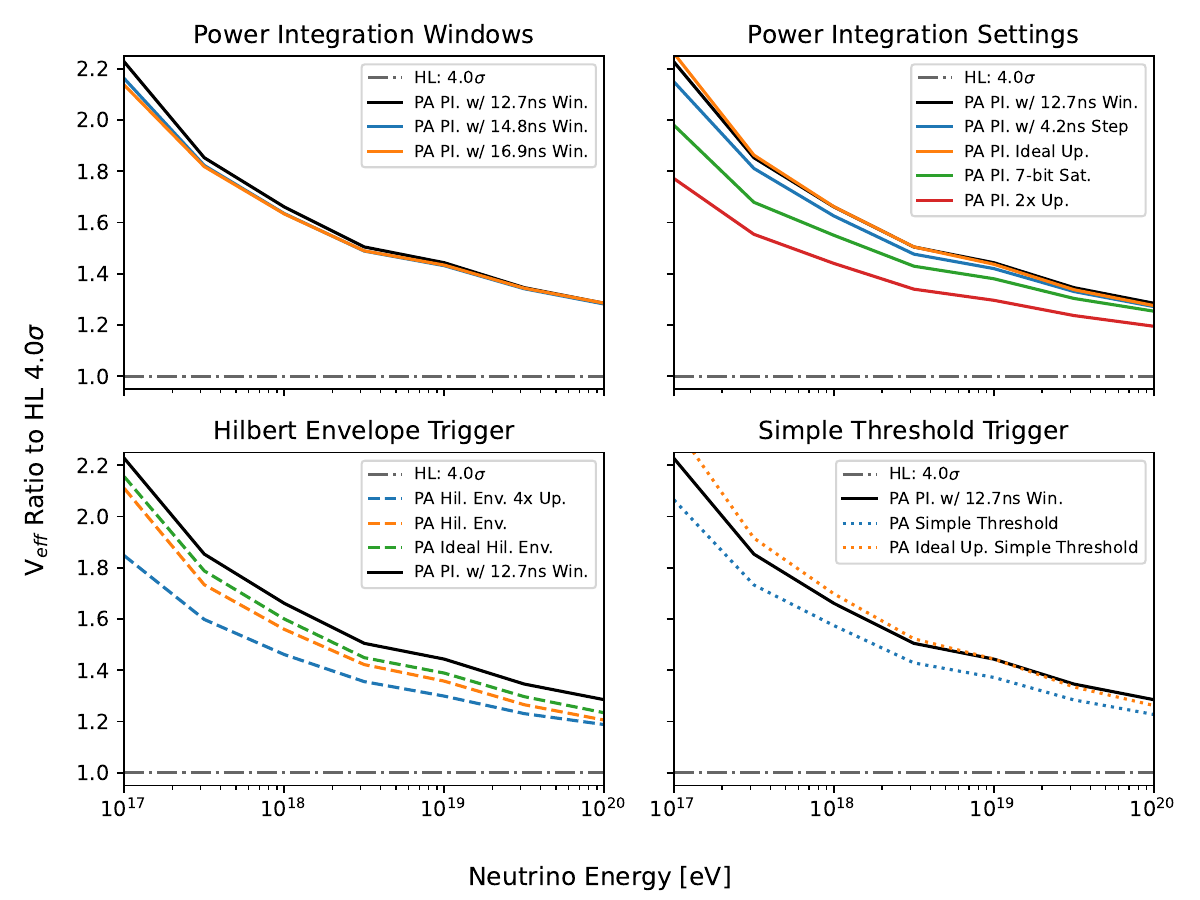}
    \caption{Comparison of various trigger methods and settings to the chosen PA trigger, ``PA PI. w/ 12.7ns Win.'', and the hi-lo trigger, ``HL: 4.0$\sigma$''. Ideal triggers are shown as a comparison to realized methods. Top left: The chosen trigger method in \ref{sec:design} with different integration windows. Top right: The integration window in \ref{sec:design} with different settings needed for a realistic implementation. Bottom left: An alternative trigger method using a Hilbert envelope. Bottom right: An alternative trigger method using a simple amplitude threshold on the beamformed voltage traces.}
    \label{fig:trig_comp}
\end{figure}

Figure~\ref{fig:trig_comp} shows the ratio of the effective volumes for different triggering schemes and settings to the hi-lo trigger and to the PA trigger presented in this section. The simulation setup is identical across all simulations and the thresholds are set to the 1\,Hz noise trigger rate. The presented trigger, ``PA PI. w/ 12.7ns Win.'', outperforms longer integration windows (top left), various upsampling and power integration settings (top right), Hilbert enveloping (bottom left), and simple threshold triggers on beamformed traces (bottom right). The chosen trigger is only overcome by perfect, but unrealizable, methods (e.g. FFT upsampling, no saturation, floating point data).

\section{Trigger validation}
\label{sec:validation}
\begin{figure}
    \centering
    \includegraphics[width=.5\textwidth]{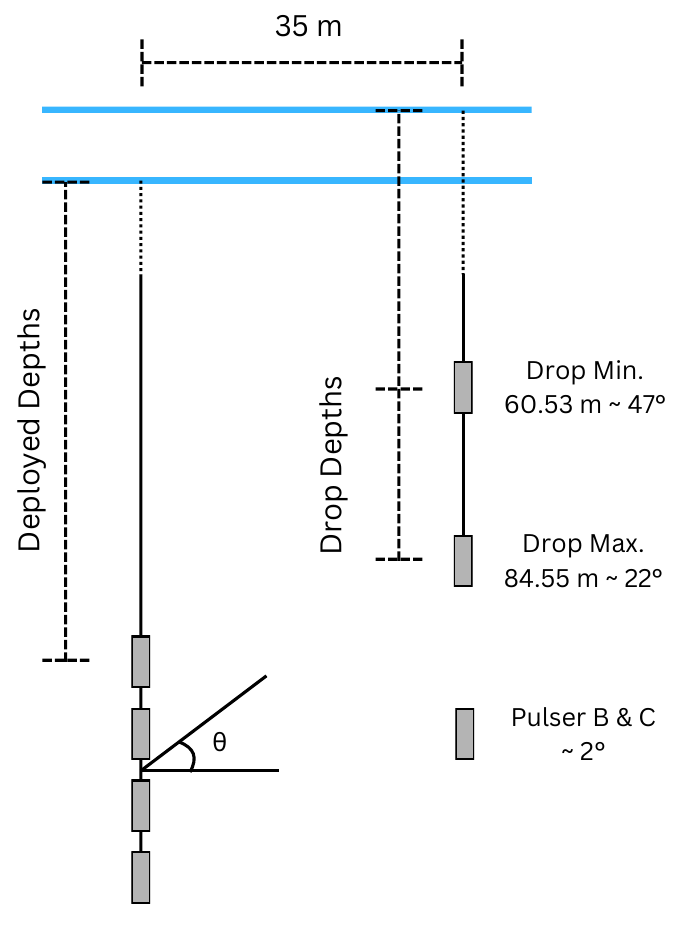}
    \caption{Diagram showing the pulser drop and nominal on-station pulser locations.}
    \label{fig:on_station_tests_diagram}
\end{figure}

% Outline methods and measurments
To validate the trigger outlined in Section~\ref{sec:design} and to compare to the previous hi-lo trigger, we performed a suite of measurements to calculate the trigger efficiencies. The first test was performed in a lab setting to isolate the firmware behavior on a reduced detector setup. The second is using the on-station calibration pulsers located on adjacent strings to the PA. The third is by performing a pulser drop, where a calibration pulser is lowered and raised in a nearby borehole to probe the trigger performance to different elevation angles. Figure~\ref{fig:on_station_tests_diagram} shows a simplified diagram illustrating the in-situ measurements, where the elevation angles are calculated from the center of the PA.

\subsection{Trigger efficiency estimation}
\label{sec:efficiency_calc}
\begin{figure}
    \centering
    \begin{minipage}{0.49\textwidth}
        \centering
        \includegraphics[width=\textwidth]{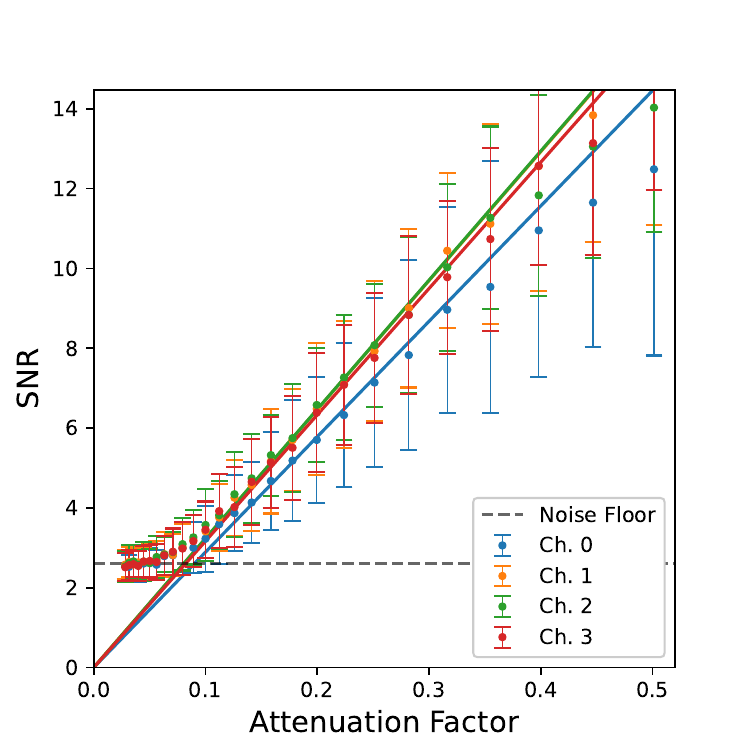}
    \end{minipage}
    \begin{minipage}{0.49\textwidth}
        \centering
        \includegraphics[width=\textwidth]{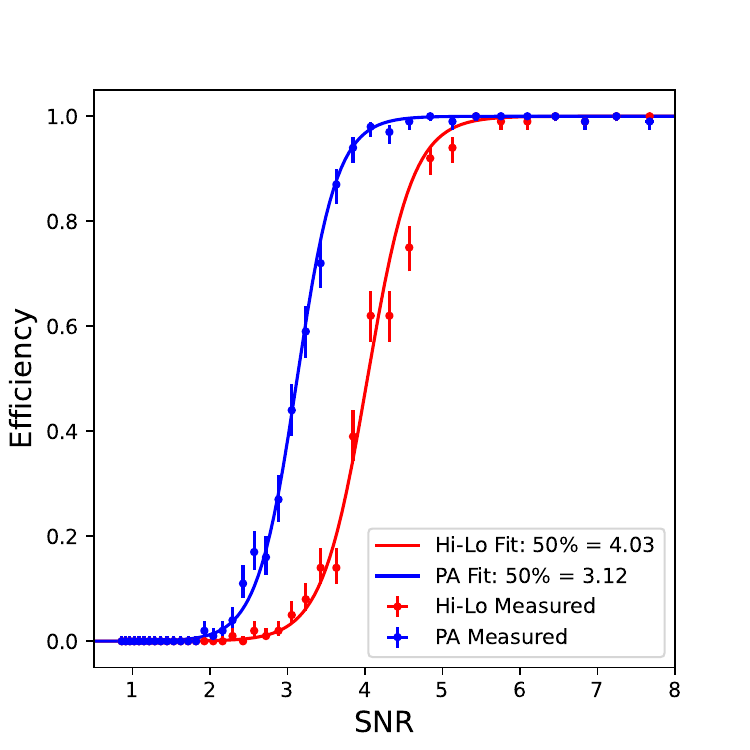}
    \end{minipage}%
    
    \caption{Example efficiency measurement using embedded calibration sources. These are made for each station and pulser. Left: Mean SNR of the as-recorded waveforms plotted against the attenuation factor setting for pulser 0 on station 14. At SNRs around 3 the event SNR saturates to the SNR of thermal noise. At SNRs above 10, the signal amplitudes begin to saturate at the maximum voltages allowed by the RFoF diodes. Right: Efficiency curve for pulser 0 on station 14. The reported SNR in the efficiency curve is the second highest SNR from the attenuation fit to match studies in Ref.~\cite{RNO-G:2025inst}, where the channel with the second highest SNR drives the 2 of 4 channel coincidence in the hi-lo trigger.}
    \label{fig:single_eff}
\end{figure}

% SNR
Because the input signal is embedded in noise we need to estimate the SNR of the input signal when the amplitudes fall below the typical thermal noise level. We use the RNO-G standard metric of $\mathrm{SNR}=\frac{V_{\mathrm{pk-pk}}}{2 V_{\mathrm{RMS}}}$ using the second-highest SNR across channels as the event's SNR to match previous studies in Ref.~\cite{RNO-G:2020rmc,RNO-G:2025inst}, where the second highest SNR drives the channel coincidence needed in the hi-lo trigger. The peak to peak voltage, $V_{\mathrm{pk-pk}}$, is calculated from the approximate location of the triggering pulses using the as-recorded waveforms. The root mean square voltage, $V_{\mathrm{RMS}}$, is calculated using the pre-pulse portion of the as-recorded traces where triggering pulses are not expected. The above-noise points with SNR$\gtrsim$5 are used to perform a linear fit of the mean SNR vs. attenuation factor in order to extrapolate SNRs to lower signal amplitudes. SNR measurements saturate to the thermal noise level of $\sim$3\,SNR. An example of this process is shown in Figure~\ref{fig:single_eff} (left). Each channel is fit individually. Error bars correspond to 1 standard deviation of individual event SNRs at an attenuation and are affected by thermal noise fluctuations and trace-to-trace variations in the sampled peak voltages. This SNR fit is performed for all station and pulser pairs.

% Efficiency
The trigger efficiency is calculated by dividing the number of triggered pulses by the total number of pulses sent, shown in Eq.~\eqref{eq:efficiency}. Pulses sent on station use a GPS-locked pulse per second (PPS). In the lab setting we use a pulse rate exceeding the possible event readout rate of $\sim$ 20\,Hz, so we instead use the trigger scalers. Scalers are triggers per time period using a set period of 1\,s. We then fit the efficiency vs. SNR with a logistic function to better estimate the 50\% efficiency point. The 50\% efficiency point represents the amplitude where 50\% of the signals trigger the system. Figure~\ref{fig:single_eff} (right) shows an example efficiency curve for pulser 0 on station 23. Both the hi-lo and PA triggers are shown.

\begin{equation}
\label{eq:efficiency}
    \varepsilon=\frac{N_{\mathrm{triggered}}}{N_{\mathrm{sent}}}
\end{equation}

\subsection{Lab testing}
\label{sec:lab_testing}

\begin{figure}
    \centering
    \includegraphics[width=.99\textwidth]{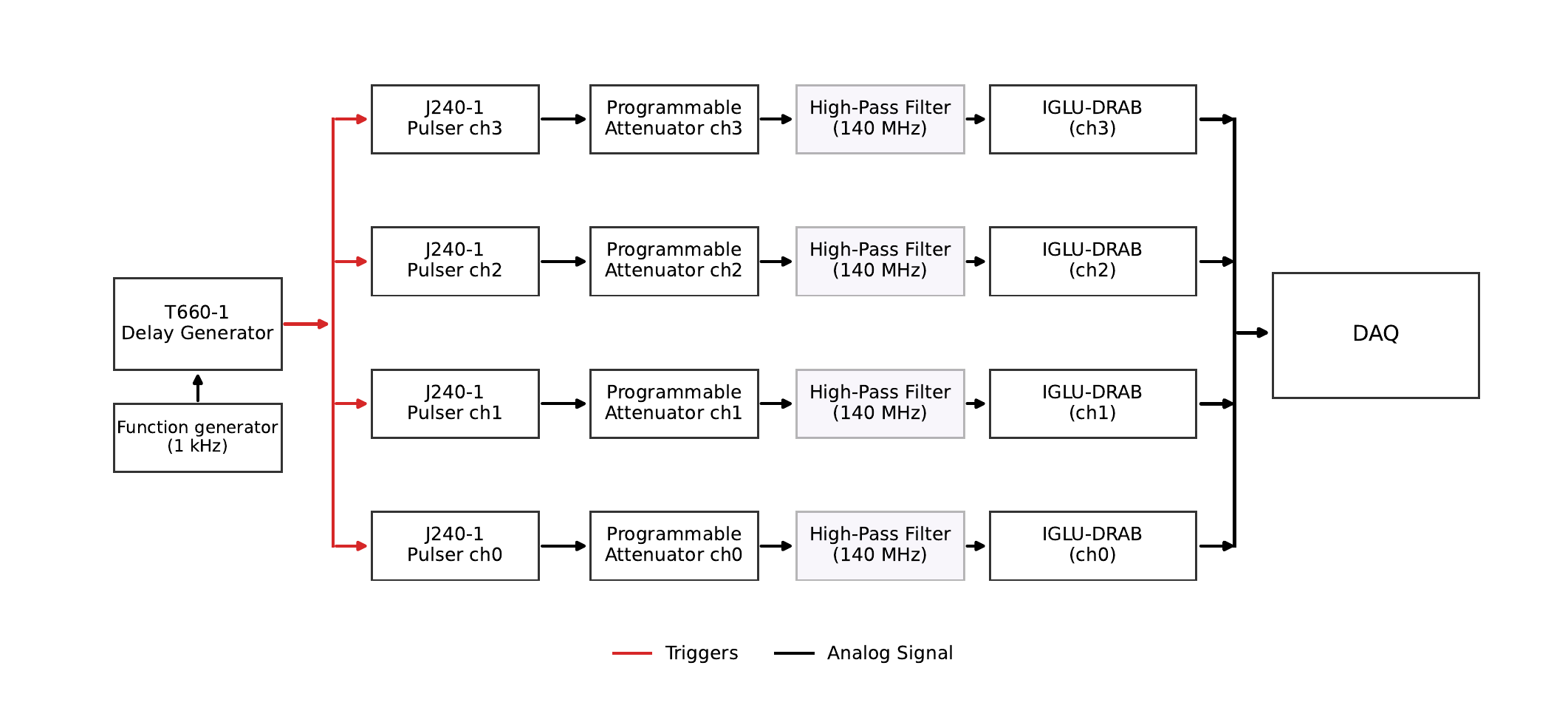}
    \caption{Block diagram for the PATT that adds adjustable delays and attenuations to test pulses.}
    \label{fig:PATT_block_diagram}
\end{figure}

We first characterize the trigger performance of the RNO-G DAQ system in a controlled lab environment using the Phased Array Trigger Tester (PATT). A schematic block diagram for the PATT is shown in Figure~\ref{fig:PATT_block_diagram}. It consists of four independently controllable fast-pulser outputs (Highland Technology J240-1), with the amplitudes controlled using a programmable attenuator board. The pulse shape is tunable through a selectable set of high-pass filters, which are used to make the unipolar pulse more similar to a bipolar neutrino-induced Askaryan pulse. Each of the four pulsers is synchronized using a delay generator, allowing small, precisely defined timing offsets to be introduced between channels to emulate signal arrival time differences at different antennas. The system was calibrated such that, for a plane wave observed at a 0\degree~view angle, all pulses arrive at the DAQ with zero relative delay. For non-zero angles, additional delays are applied on top of this calibration baseline. The scans were performed in 0.5\degree~elevation angle increments from -60\degree~to 60\degree~and over an attenuation range spanning the full transition from 0\%--100\% efficiency.

The output of the PATT is propagated through the full IGLU-DRAB signal chain and subsequently through the FLOWER system, but does not include the angle dependence or the dispersion of the antenna response. Trigger thresholds on the FLOWER are determined before sending pulses by enforcing a 1\,Hz trigger rate on pure thermal noise. The PATT operates at a fixed input pulse rate of 1\,kHz, allowing for efficient parameter scans over different elevation angles. The ratio of the measured trigger rate to the input pulse rate directly defines the efficiency. Both the hi-lo trigger and the PA trigger are scanned.

\begin{figure}[htbp]
    \centering
    \includegraphics[width=.8\textwidth]{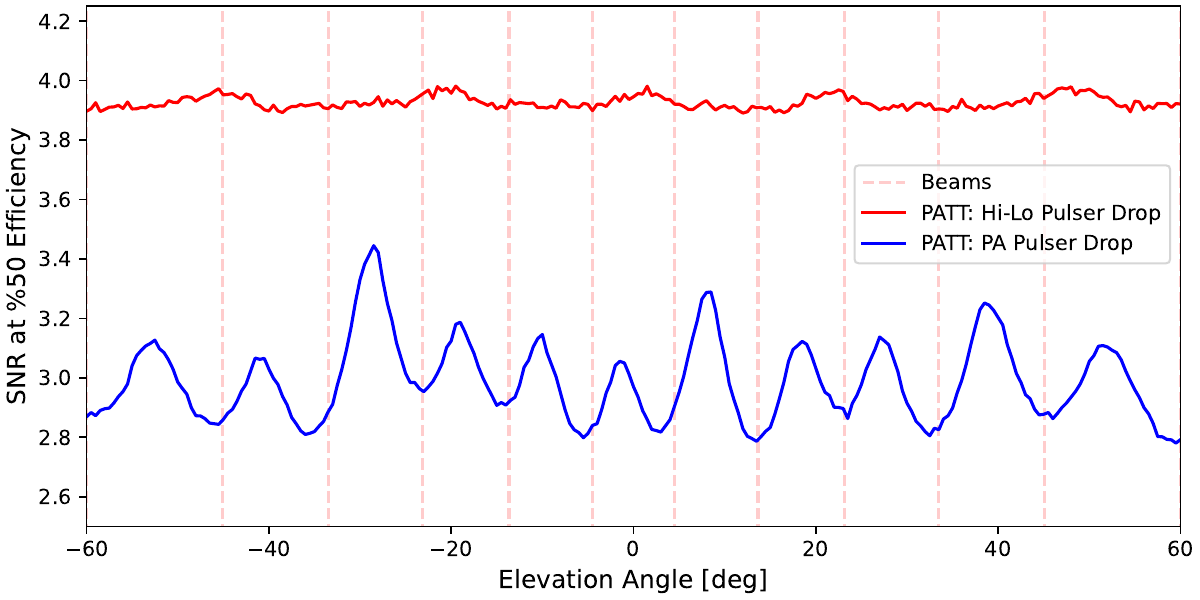}
    \caption{Measured trigger efficiency scans using expected signal arrival times for station 11 geometry with a plane wave approximation. Generated pulses and noise are passed through a real RF signal chain but do not contain the imprint of the antenna response.}
    \label{fig:patt_testing}
\end{figure}

The lab scans provide measurements of the performance of both the hi-lo trigger and the PA trigger across its 12 beams. Figure~\ref{fig:patt_testing} summarizes these results. As expected, the PA trigger demonstrates substantially improved performance, achieving 50\% efficiency at lower SNR values relative to the hi-lo trigger. Across all angles the average SNR at 50\% efficiency is approximately $3.9\,\sigma$ for the hi-lo trigger and $3.0\,\sigma$ for the PA trigger. We note that the main difference between lab and on-station measurements is that the former do not take into account the effects of the antenna response which affects the amplitude of signals received in each channel and the ice propagation which changes signal arrival times due to the changing index of refraction.

\subsection{On-station testing}
\label{sec:on-station}

% In-situ Pulsers and Pulser drop
To measure the performance of the triggers in the full detector we use the in-situ pulsers. The PA has been deployed to the current 8 stations since the 2025 season and each station has 2 pulsers permanently located on the helper strings. Figure~1 of Ref.~\cite{RNO-G:2025inst} shows a diagram of the typical station geometry. Pulses are generated and attenuated on a dedicated pulser board in the DAQ box. Attenuations range from 0 to -31.5 dB in 0.5 dB steps. During a calibration run, the station is configured like a normal physics run, but with the calibration pulser enabled to fire on the GPS PPS. Pulses are sent at each attenuation for 100 seconds. 

\begin{figure}
    \centering
    \includegraphics[width=.7\textwidth]{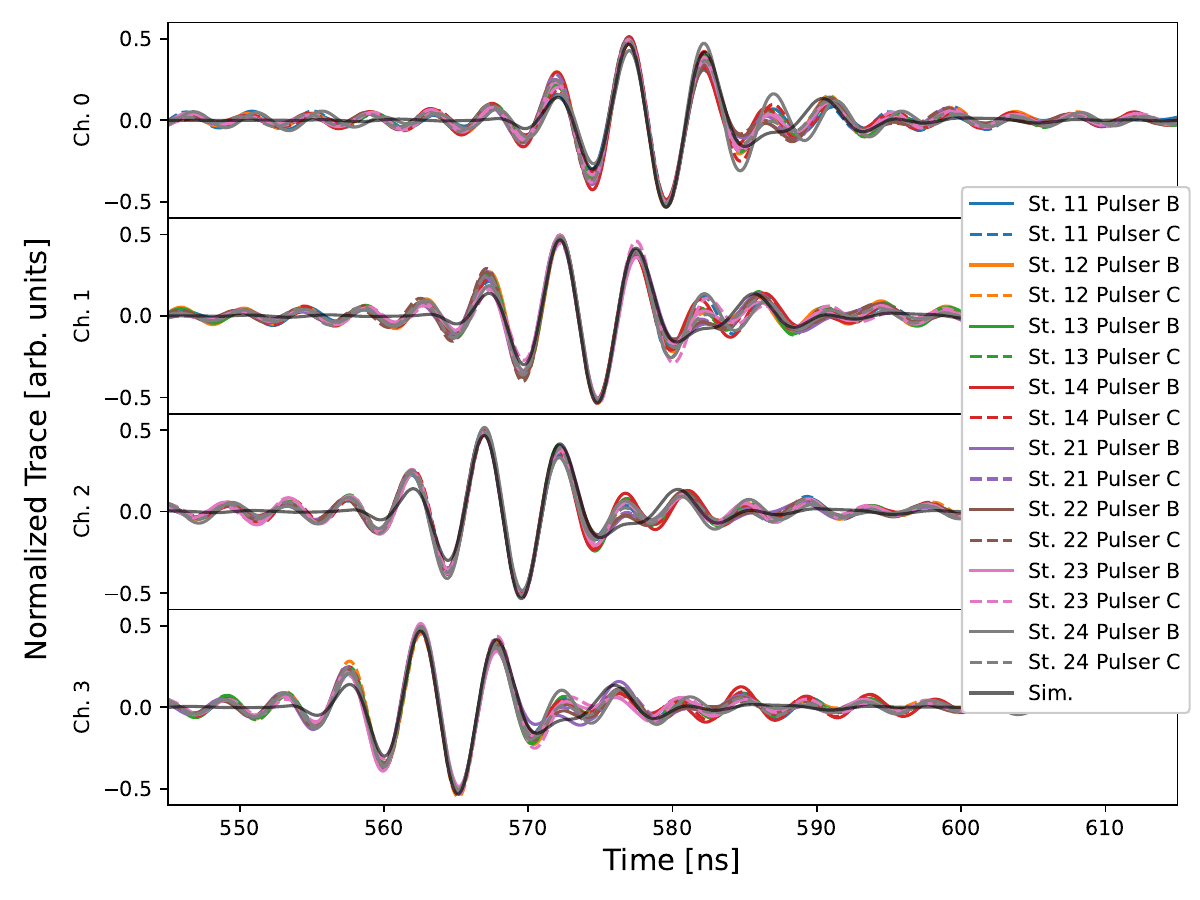}
    \caption{Average calibration pulses in the trigger band as recorded in each station for each on-station pulser. The simulated pulser signal is shown as well.}
    \label{fig:on_station_pulsers}
\end{figure}

Using recorded pulses on all stations, we first check that they agree across stations and to the simulated model. Figure~\ref{fig:on_station_pulsers} shows the averaged calibration pulse for each station and each pulser compared to the simulated equivalent. For comparison the signals are aligned in time for each channel. Time delays between channels are not shown. Good agreement between real recorded signals and simulated pulses suggest that the calibration pulser is a good proxy to an on-cone Askaryan pulse as seen in Figure~\ref{fig:nu_traces}.

\begin{figure}
    \centering
    \includegraphics[width=1\textwidth]{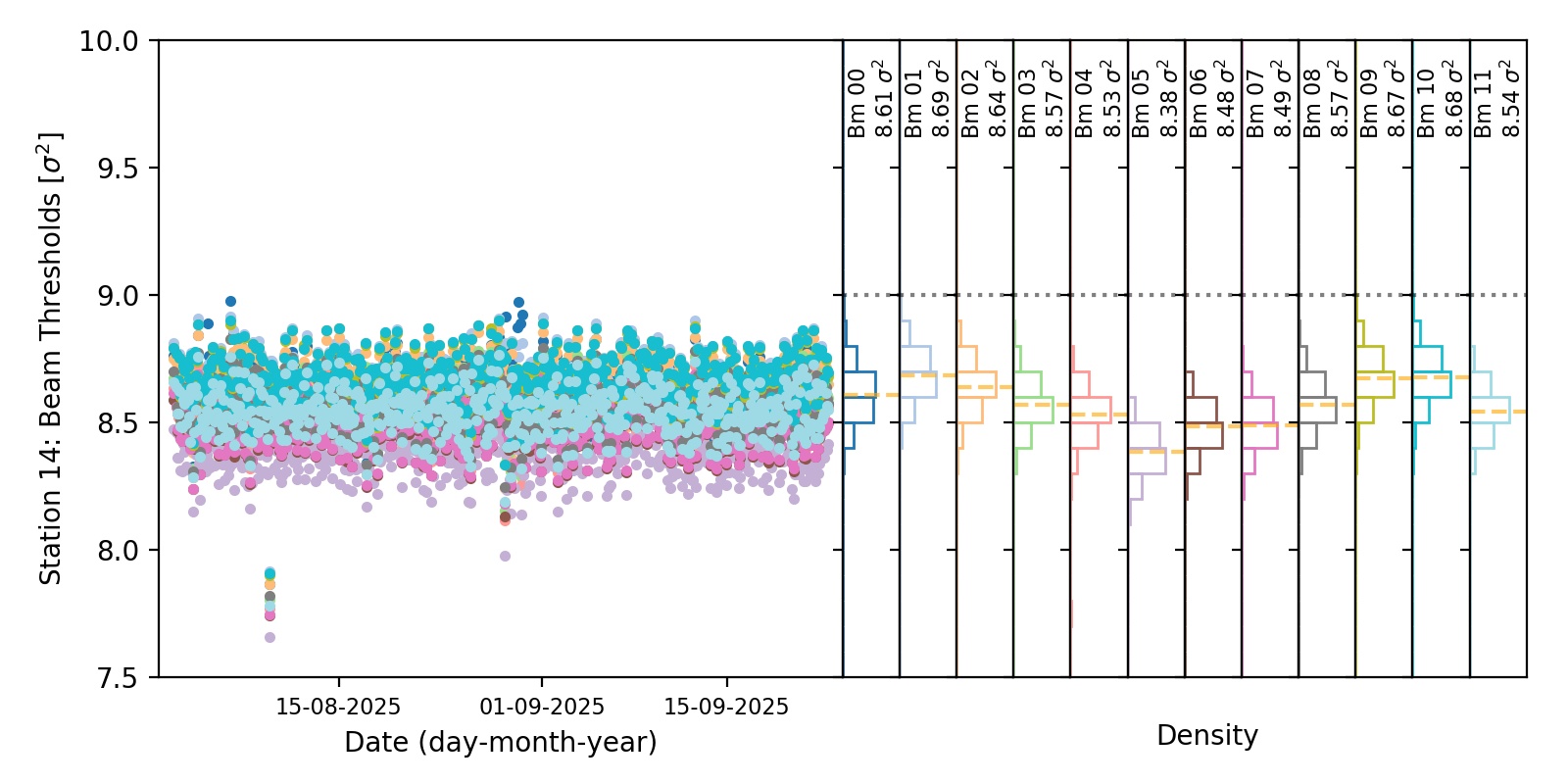}
    \caption{Beam thresholds starting in August 1st, 2025 until the end of data collection in October 2025 on station 14. Stations 11, 12, 13, 14, and 23 have similar signal chains whose thresholds are typically below the predicted level. Stations 21, 22, and 24 have similar signal chains used during 2021-2024 whose thresholds match simulated levels. Left: Median run thresholds in units of a factor of the coherently summed RMS, $\sigma^2$, over time. Deviations above the average are caused by background contamination during a run, and deviations below are due to contamination during measurement of the $V_{\mathrm{RMS}}$ preceding a run to convert the raw ADU$^2$ threshold to factors of $\sigma^2$. Right: Histograms of each beam's median run threshold with the average threshold calculated compared to the simulation derived threshold (dashed horizontal line).}
    \label{fig:station_thresholds}
\end{figure}

% Thresholds
Trigger thresholds are adjusted using a servo method (found in Ref.~\cite{RNO-G:2025inst}, but is explained here). Two sets of thresholds are used. The servo thresholds are set lower than the trigger thresholds so that they trigger on the noise on the order of 1000 times per second. The servo rate is called the scaler goal. The trigger thresholds are calculated using the servo threshold divided by a set servo fraction. A sample of on station trigger thresholds can be seen in Figure~\ref{fig:station_thresholds}. Each beam has its own independent threshold. These thresholds are shown as factors to the square of the coherently summed RMS ($\sigma^2=\sum_{ch} \sigma_i^2$ where $\sigma$ is the coherently summed RMS and $\sigma_i$ is the RMS of each channel). Each beam has a target servo rate which causes thresholds to vary across the beams. Beam thresholds are lowest at boresight of the antennas, beams 5 and 6, and maximum for beams located near $\pm50$\degree, beams 1 and 10. We note that there are variable thresholds across the beams. The thresholds scale with the noise power in the beamformed traces in RADIANT-captured software trigger data, suggesting low-level correlated noise in the environment.

With trigger thresholds established for each station, we compute the resulting 50\% trigger efficiency for each station and on-station pulser, shown in Table~\ref{tab:table_efficiencies}. Table~\ref{tab:table_efficiencies} shows the 50\% efficiency point for all on-station pulser and station combinations. For nominal stations (12, 14, 23, and 24), the average efficiency of the hi-lo trigger and the PA trigger are 4.1 and 3.1, respectively. The PA trigger is a significant improvement to the hi-lo trigger.

\begin{table}[bp]
    \centering
        \caption{Trigger efficiencies for each station and each in-situ pulser for the hi-lo trigger (2021-2024) and the trigger presented here (2025+). Nominal stations (12, 14, 23, 24) are highlighted in gray. We caution reading into station specific differences, particularly results for stations 11, 13, 21, 22, in gauging a typical station performance as these either have known concerns with the pulser, a bad channel, or unknown deviation from the expected performance.}
    \begin{tabular}{|c|c|c|c|c|}
         \multicolumn{5}{c}{50\% Trigger Efficiencies to In-Situ Pulsers [SNR]}\\
    \hline
         Station&  Pulser B: Hi-Lo &  Pulser B: PA&  Pulser C: Hi-Lo& Pulser C: PA\\
    \hline
         11& 4.38 $\pm$ 0.08 & 2.42 $\pm$ 0.07 & 4.51 $\pm$ 0.07 & 2.61 $\pm$ 0.04\\
         \rowcolor{lightgray} 
         12& 4.07 $\pm$ 0.05 & 2.99 $\pm$ 0.04 & 3.88 $\pm$ 0.04 & 3.04 $\pm$ 0.08\\
         13& 4.29 $\pm$ 0.05 & 3.54 $\pm$ 0.03 & 3.97 $\pm$ 0.04 & 3.44 $\pm$ 0.02\\
         \rowcolor{lightgray} 
         14& 4.03 $\pm$ 0.03 & 3.12 $\pm$ 0.02 & 3.75 $\pm$ 0.03 & 3.04 $\pm$ 0.04\\
         21& 3.24 $\pm$ 0.03 & 2.45 $\pm$ 0.03 & 3.03 $\pm$ 0.04 & 2.46 $\pm$ 0.04\\
         22& 4.83 $\pm$ 0.07 & 3.19 $\pm$ 0.04 & 4.69 $\pm$ 0.08 & 3.47 $\pm$ 0.04\\
         \rowcolor{lightgray}
         23& 4.03 $\pm$ 0.09 & 3.02 $\pm$ 0.07 & 4.00 $\pm$ 0.10 & 2.99 $\pm$ 0.09\\
         \rowcolor{lightgray}
         24& 3.80 $\pm$ 0.02 & 3.05 $\pm$ 0.02 & 3.72 $\pm$ 0.03 & 2.75 $\pm$ 0.03\\
    \hline
    \end{tabular}

    \label{tab:table_efficiencies}
\end{table}

We attribute large variations in measured efficiencies to a few factors. For station 11, the calibration pulser can fire several times per GPS pulse. The hi-lo trigger is not as affected, but the PA trigger is, lowering the 50\% point by at least 0.5\,SNR. For station 13, one channel exhibits much lower gain and shows signs of a bad connection to the antenna, effectively leaving only three working channels and degrading trigger performance. For stations 21 and 22, both stations were installed in the first year of deployment, where a board-level fix was required for the DRAB boards in the field, and are outliers in this analysis that may be impacted by this effect. Their simulated beam thresholds match the measured thresholds well, but the 50\% point for both the hi-lo and PA triggers does not match expectations. The positions of the pulsers are not identical throughout the array. Drill issues in the earlier seasons caused some of the helper strings to be shallower compared to the PA depths, causing differences in parts of the beam pattern observed. Finally, each station can run at slightly different trigger rates and thresholds. Scaler goals are tuned such that the trigger rate is in the range of 0.9-1.2\,Hz, but background noise levels can cause the thresholds to increase.

The second in-situ test we perform is a pulser drop. During a pulser drop the calibration pulser is moved through an adjacent borehole in order to probe the PA to various elevation angles. This test was performed on station 14 where there is a spare borehole $\sim$35m away from the PA. An example of this for other stations can be found in Ref.~\cite{RNO-G:2025inst} which measured the antenna beam pattern, ice behavior, and reconstruction algorithms. Given limited deployment time and the need to probe several different attenuations and depths, we use much shorter scan times compared to the on-station pulser measurements. Only 7 attenuations are scanned at each depth with the SNRs surrounding the 50\% efficiency point. At each attenuation, 60 waveforms are recorded to map the SNRs and short 60-second runs are taken with the different triggers running to measure the trigger efficiency.

\begin{figure}[htbp]
    \centering
    \includegraphics[width=.7\textwidth]{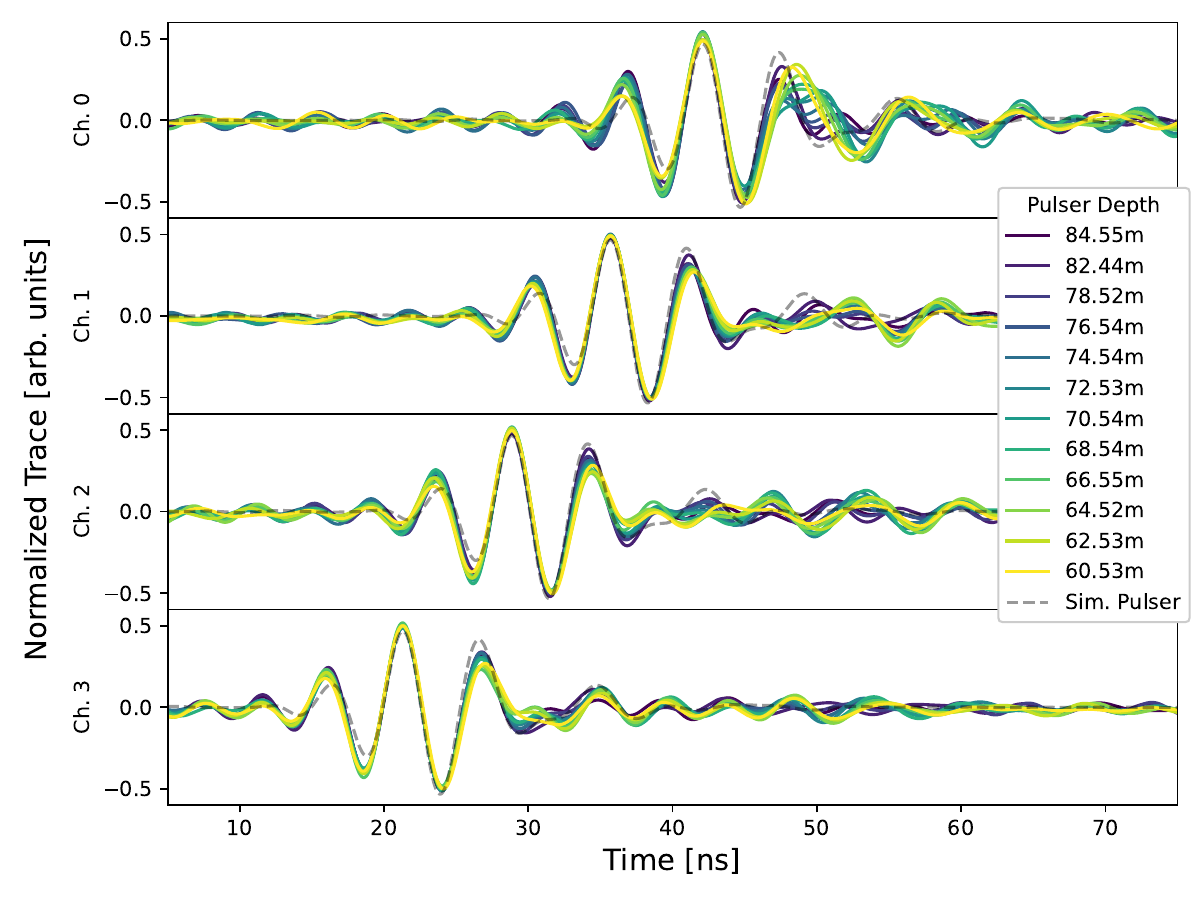}
    \caption{Average pulser waveforms at each depth during the pulser drop. The signal shape is not expected to change greatly from the phase response of the antenna pattern. Signals in individual channels are aligned in time for shape comparison.}
    \label{fig:pulser_drop_waves}
\end{figure}

Figure~\ref{fig:pulser_drop_waves} shows the average pulses at each depth. The pulse shape is relatively unchanged as is expected from Vpol simulations. Although not shown, the amplitude does change. However, events are normalized to the noise through the SNR, which results in trigger efficiencies that are invariant to the voltage amplitudes.

\begin{figure}
    \centering
    \includegraphics[width=1\textwidth]{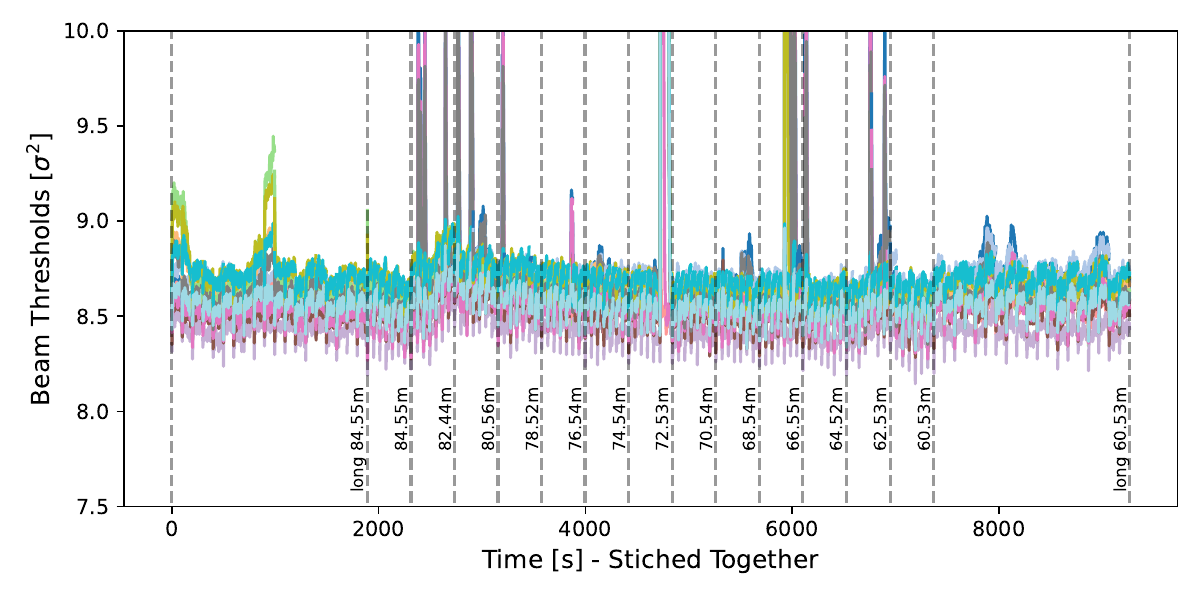}
    \caption{Beam thresholds over the span of the pulser drop. Efficiency data points were removed for thresholds over 10$\sigma^2$.}
    \label{fig:drop_thresholds}
\end{figure}

To compare the short time period of this measurement to the long-term operations, Figure~\ref{fig:drop_thresholds} shows the beam thresholds during the pulser drop. Separate runs at each depth are stitched together for clarity. The thresholds are relatively stable with the exception of a few prominent spikes, which are removed from the efficiency calculations. These thresholds are comparable to those in Figure~\ref{fig:station_thresholds}.

\begin{figure}
    \centering
    \includegraphics[width=1\textwidth]{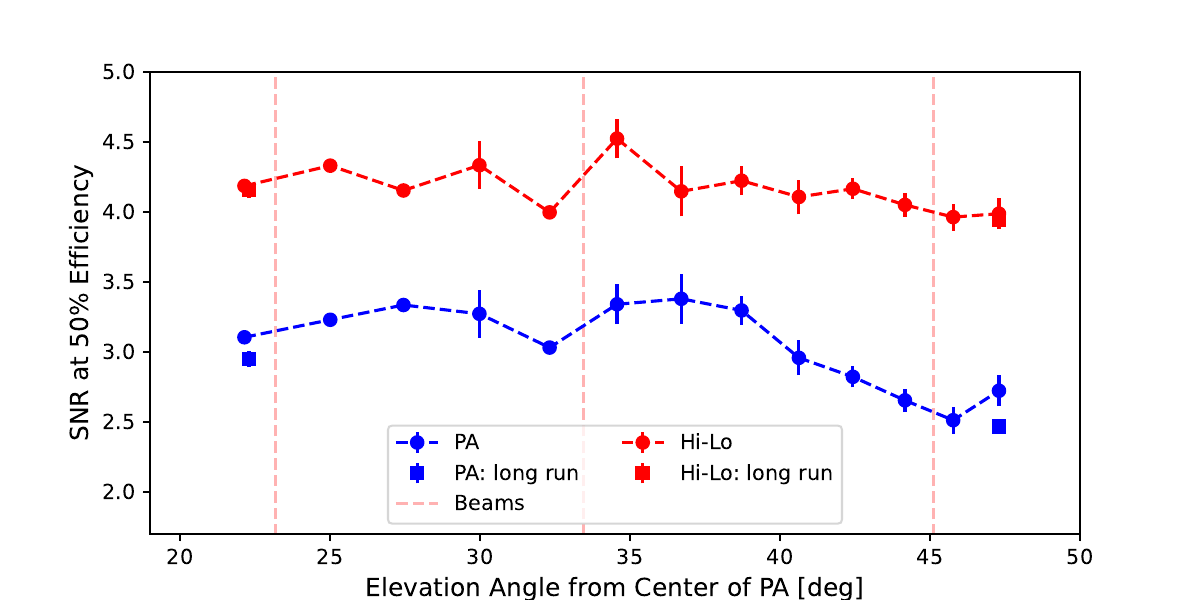}
    \caption{50\% trigger efficiency points during the 2025 pulser drop measurements. We highlight the improvement of the new PA trigger (blue) to the previously running hi-lo trigger (red) over a limited set of elevation angles compared to the full beam pattern of the trigger. ``Long run'' indicates longer runs and waveform collection with better statistics performed at the beginning and end of the pulser drop.}
    \label{fig:pulser_drop_eff}
\end{figure}

The pulser drop helps to understand the trigger response to different incident elevation angle at the PA. Figure~\ref{fig:pulser_drop_eff} shows the efficiency as a function of elevation angle to the center of the PA channels. We see good improvement with the PA trigger across all tested elevation angles, consistent with the lab scans in Figure~\ref{fig:patt_testing}. Longer runs are performed at the start and end points of the scan. The 50\% points for the hi-lo trigger agree, but do not agree statistically for the PA trigger, although they are still within 10\% of each other.

\begin{figure}

    \centering
    
    \begin{minipage}{1\textwidth}
        \centering
        \includegraphics[width=.8\textwidth]{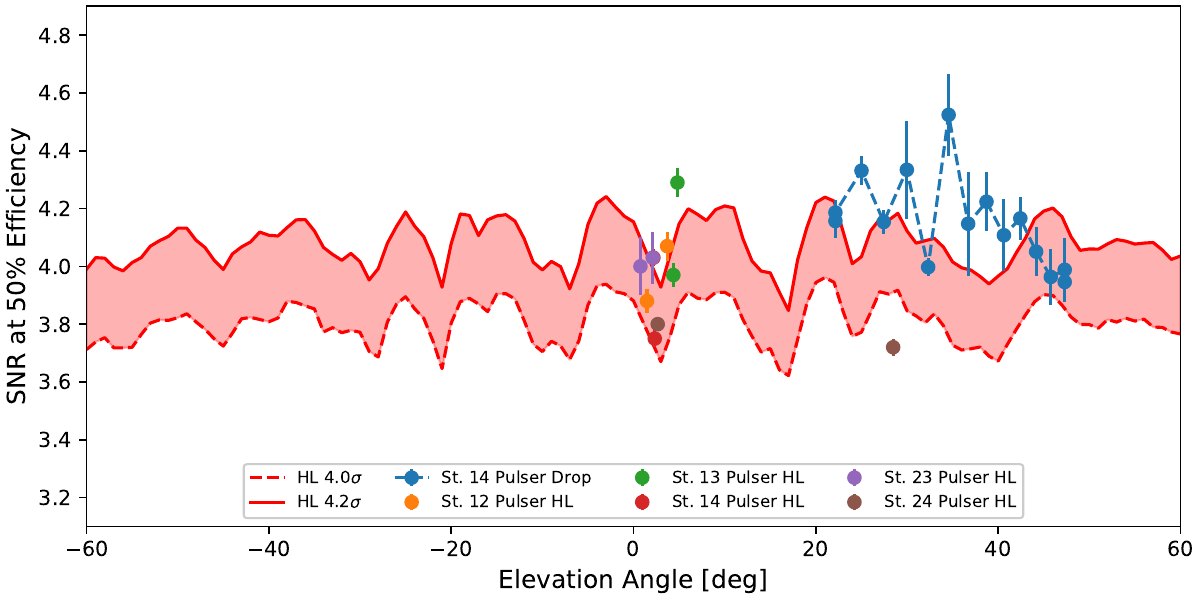}
        \vspace{.3cm}
    \end{minipage}

    \begin{minipage}{1\textwidth}
        \centering
        \includegraphics[width=.8\textwidth]{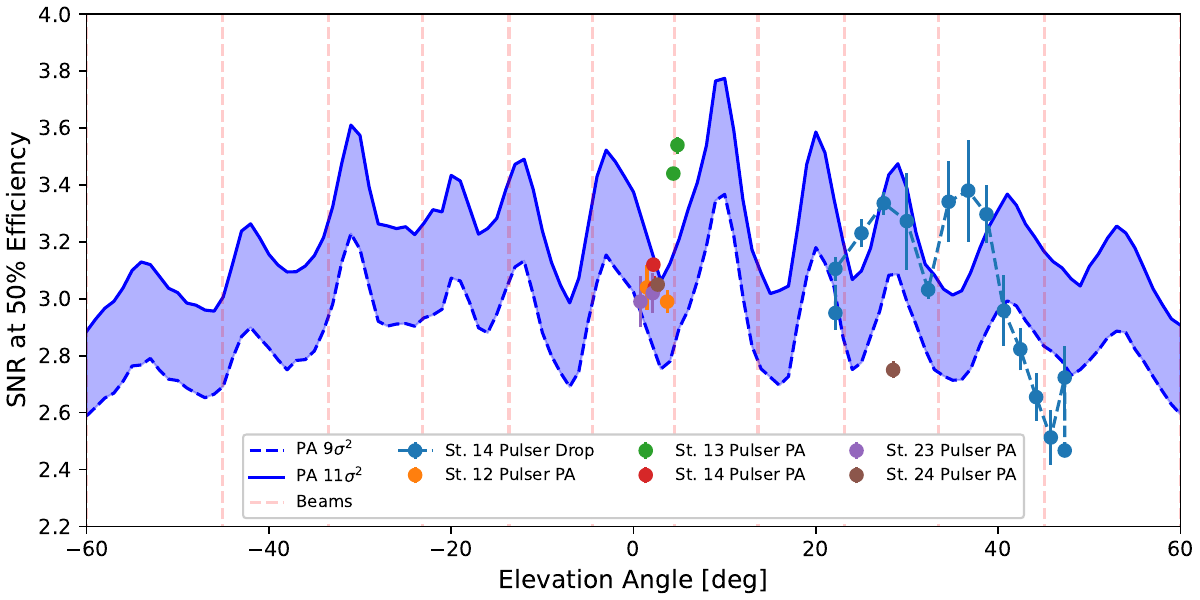}
    \end{minipage}

    \caption{Simulated and measured calibration pulser 50\% efficiency points for the hi-lo trigger (top) and the PA trigger (bottom). Outlier points greater than 1 SNR from expected are not shown. Highlighted bands are shown with the lower limit set by simulation derived thresholds, and an upper limit set by higher thresholds that match the observed data. Simulations are performed on station 23 lab-measured signal chain responses, station geometry, and measured voltage pulse of the calibration pulser.}
    \label{fig:full_pulser_drop_eff}
\end{figure}

The results of the simulated pulser drops are shown alongside measured station data in Figure~\ref{fig:full_pulser_drop_eff}. Station 23 is used as the representative station for simulations. Figure~\ref{fig:full_pulser_drop_eff} (top) shows the hi-lo trigger. Figure~\ref{fig:full_pulser_drop_eff} (bottom) shows the PA trigger. The hi-lo trigger is mostly independent of elevation angle, as expected. The PA trigger achieves a lower overall threshold and an approximate alignment of the minima in the 50\% efficiency point with the chosen beam locations until the drop-off above $\sim$\ang{38}. The abnormal beam pattern may be due to the sparse data collected that was limited to only a single attempt, to issues with our current ice model, or to the plane wave approximation in beamforming.
 
We note that using the simulation-derived thresholds overestimates the performance of the detector to our calibration pulsers for both triggers. To reconcile this discrepancy when estimating the effective volumes in Section~\ref{sec:effective_volume}, we adjust the thresholds such that the simulations match the measured efficiencies. The pulser-matched threshold for the hi-lo trigger is 4.2~$\sigma$ and for the PA trigger is 11~$\sigma^2$, whereas simulation derived thresholds are 4.0~$\sigma$ and 9.0~$\sigma^2$. The 50\% efficiency point for both sets of curves is shown in Figure~\ref{fig:full_pulser_drop_eff}.

\section{Effective volume ratio}
\label{sec:effective_volume}

Finally, we calculate the ratios of the single-station effective volumes using the hi-lo trigger and the PA trigger to estimate the detector's improvement to neutrinos. Simulations are also performed with the \texttt{NuRadioMC}~\cite{Glaser:2019cws,Glaser_2019} framework, discussed in Sec.~\ref{sec:nu_sims} and reiterated here for the specific models used.

% the detector
The detector is modeled with realistic and accurate trigger models, verified by RTL simulations and the measurements in Section~\ref{sec:validation}. We also use the measured signal chains of the IGLU, DRAB, and digitizer boards, as measured in the lab before station deployment. Updated antenna models are used, and validated with in-situ comparisons as well~\cite{RNO-G:2025inst}. We use station 23's description as the representative station for this simulation too. 

% other sim parameters

The ice attenuation model, GL3, is sourced from attenuation length ice measurements at Summit Station~\cite{GL3} and similarly the index of refraction model is a single-exponential model from data in Ref.~\cite{cosmin_ice}. 
Primary neutrino events are generated using the CTW cross sections and inelasticities~\cite{ctw}, which determines charged current and neutral current interaction fractions and deposited shower energies, respectively. Subshowers from propagating secondary particles~\cite{rnog_sec} are generated with stochastic depositions and tau decays in \texttt{PROPOSAL}~\cite{proposal}. Askaryan pulses are generated with the semi-analytic model of radio emission from particle showers, ARZ2020~\cite{ARZ2020}.

\begin{figure}
    \centering
    \includegraphics[width=.9\textwidth]{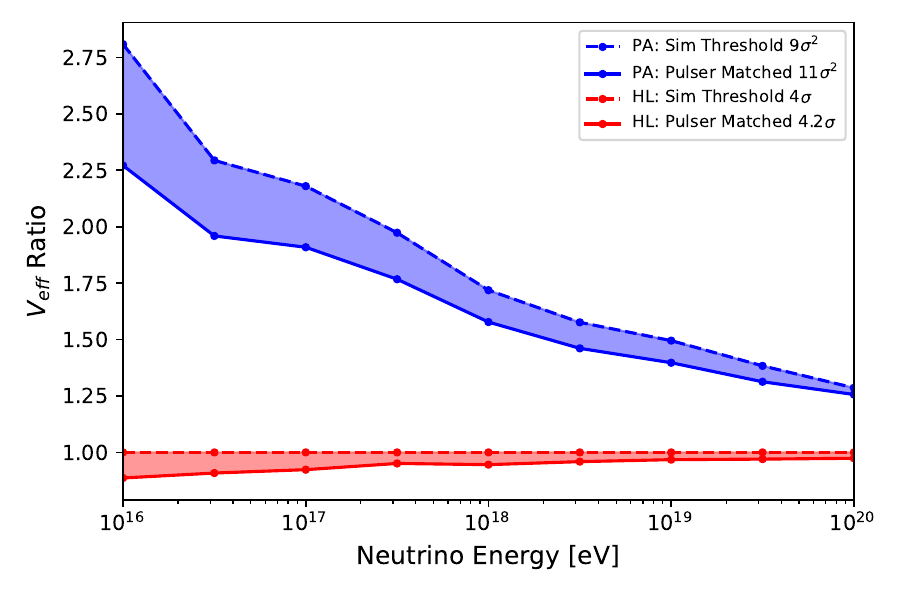}
    \caption{Ratio of single station effective volume using the simulation and data derived thresholds of the presented PA trigger, and the previous hi-lo trigger.}
    \label{fig:eff_volume}
\end{figure}

% results
Ratios of the single-station effective volumes are shown in Figure~\ref{fig:eff_volume} using the simulation-derived trigger thresholds, 4$\sigma$ and 9$\sigma^2$, and the data-driven trigger thresholds, $4.2\sigma$ and 11$\sigma^2$, for the hi-lo and PA triggers. At 100\,PeV and below, the PA trigger improves by at least a factor of $\sim$2. The improvement decreases for higher energies, but stays upwards of a factor of $\sim$1.3. The band for each trigger accounts for discrepancy in performance between simulation and data and is at most 10\% different.

Better-performing triggers are more sensitive to weaker Askaryan signals. The geometry of an event, vertex position and neutrino direction, can directly affect how strong the received signal is. This includes particle showers that are more distant, Askaryan signals with a polarization that tilts further away from Vpol, or Askaryan signals that are seen further off-cone. Better-performing triggers are more impactful in the lower energy region, where the expected neutrino flux is higher. The inelasticity determines the energy of the resulting particle shower and thereby the strength of the Askaryan signal. The distribution of inelasticities is more heavily weighted to low energy transfer from neutrinos to the ice. For near-threshold neutrinos, a better trigger improves the detector volume by increasing the number of lower inelasticity showers that can be detected, while at high energies most of the volume improvement is attributed to the geometrical effect.

We expect future improvements on upcoming stations as state-of-the-art electronics become more resource-efficient at a given power budget, allowing us to implement more advanced triggers, or second-stage triggers, that filter thermal noise triggered events~\cite{rnog:2023_ml_trigger, arianna_2022_ml} at the FPGA level.

\section{Conclusion}
RNO-G is a radio array aimed at detecting the Askaryan radio emission of UHE neutrinos in the Greenland ice sheet. This work presents a new phased-array (PA) trigger that improves the sensitivity of the detector by a factor of 1.3 at the highest energies and by at least a factor of 2 at energies below 0.1\,EeV compared to the previously running high-and-low voltage-threshold (hi-lo) trigger. Through upsampling, delay-and-sum beamforming, and power integration, this trigger better targets the expected neutrino-induced Askaryan emission. This trigger has been running on each of the 8 currently deployed stations since the 2025 data collection season. Both the new trigger and the prior trigger were characterized using measurements in the lab and at the stations. This work marks a significant step forward for RNO-G, realizing a digital phased-array trigger that had previously existed only in simulation.

\section{Acknowledgments}
We are thankful to the support staff at Summit Station for making RNO-G possible. We also acknowledge our colleagues from the British Antarctic Survey for building and operating the BigRAID drill for our project.

We would like to acknowledge our home institutions and funding agencies for supporting the RNO-G work; in particular the Belgian Funds for Scientific Research (FRS-FNRS and FWO) and the FWO programme for International Research Infrastructure (IRI), the National Science Foundation (NSF Award IDs 2411590, 2514262, 2514201, 2514208, 2514206, and collaborative awards 2310122 through 2310129), and the IceCube EPSCoR Initiative (Award ID 2019597), the Helmholtz Association, the German Research Foundation (DFG, Grant 556092823), the Swedish Research Council (VR, Grant 2023-00156), the University of Chicago Research Computing Center, and the European Union under the European Union's Horizon 2020 research and innovation programme (ERC, Pro-RNO-G No 101115122 and NuRadioOpt No 101116890). 

\bibliographystyle{JHEP}
\bibliography{Bibliography}

\clearpage

\end{document}